\documentclass[%
 aip,
 amsmath,amssymb,
 reprint,%
]{revtex4-1}

\draft 

\usepackage[latin1]{inputenc}
\usepackage{blindtext}
\usepackage{centernot}
\usepackage{graphicx}
\usepackage{amsmath,bbold}
\usepackage{times}
\usepackage{amssymb}
\usepackage{mathrsfs}
\usepackage{chemarr}
\usepackage{color}
\usepackage{url}
\usepackage{version}
\usepackage[hidelinks]{hyperref}
\usepackage{mwe,tikz}
\usepackage[percent]{overpic}
\usepackage{bm}
\usepackage[export]{adjustbox}
\definecolor{linkcolor}{rgb}{0,0,0.6} 
\usepackage{lipsum}

\newcommand{\niceVec}[2]{\vec{#1}^{\,#2}}
\newcommand{\mat}[1]{{\boldsymbol #1}}

\definecolor{safeGreen}{rgb}{0.0, 0.5, 0.0}              

\newcommand{\ERR}{E_\text{RR}}
\newcommand{\ERB}{E_\text{RB}}
\newcommand{\EBB}{E_\text{BB}}

\newcommand{\EHH}{E_\text{HH}}
\newcommand{\EHP}{E_\text{HP}}

\newcommand{\ENH}{E_\text{NH}}
\newcommand{\ENP}{E_\text{NP}}

\newcommand{\TF}{T_\text{f}}
\newcommand{\tc}{t_c}

\newcommand{\kt}{k_\text{B}T}

\newcommand{\tzero}{t_0}
\newcommand{\lzero}{l_0}
\newcommand{\Lzero}{L_0}

\newcommand{\np}{n_\text{P}}
\newcommand{\nh}{n_\text{H}}
\newcommand{\IH}{I_\text{H}}
\newcommand{\bhh}{b_\text{HH}}
\newcommand{\bhp}{b_\text{HP}}
\newcommand{\Rnp}{R_\text{np}}
\newcommand{\Rcone}{R_\text{cone}}
\newcommand{\se}{\sigma_\text{M}}

\begin{document}

\title{Optimization of Non-Equilibrium Self-Assembly Protocols Using Markov State Models}



\author{Anthony Trubiano}
\email[]{trubiano@brandeis.edu}
\author{Michael F. Hagan}
\email[]{hagan@brandeis.edu}
\affiliation{Martin Fisher School of Physics, Brandeis University, Waltham, Massachusetts 02454, USA}

\date{\today}

\begin{abstract}

The promise of self-assembly to enable bottom-up formation of materials with prescribed architectures and functions has driven intensive efforts to uncover rational design principles for maximizing the yield of a target structure. Yet, despite many successful examples of self-assembly, ensuring kinetic accessibility of the target structure remains an unsolved problem in many systems. In particular, long-lived kinetic traps can result in assembly times that vastly exceed experimentally accessible timescales. One proposed solution is to design non-equilibrium assembly protocols in which system parameters change over time to avoid such kinetic traps. 
Here, we develop a framework to combine Markov state model (MSM) analysis with optimal control theory to compute a time-dependent protocol that maximizes the yield of the target structure at a finite-time. We present an adjoint-based gradient descent method that, in conjunction with MSMs for a system as a function of its control parameters,  enables efficiently optimizing the assembly protocol. We also describe an interpolation approach to significantly reduce the number of simulations required to construct the MSMs. 
We demonstrate our approach on two examples; a simple semi-analytic model for the folding of a polymer of colloidal particles, and a more complex model for capsid assembly. Our results show that optimizing time-dependent protocols can achieve significant improvements in yields of selected structures, including equilibrium free energy minima, long-lived metastable structures, and transient states. 

\end{abstract}


\maketitle 

\section{Introduction}
Designing building blocks that are pre-programmed to self-assemble into a target structure has enabled creating microscopic and nano-scale materials with desirable properties and promising applications \cite{Garg2015, Beija2012, Ebbens2016, Mallory2018, Fan2011, Huh2020, Ke2012}. However, achieving high yields and selective assembly of the target structures remains a critical unsolved problem in most systems. Thus, there has been intense research aimed at discovering the governing principles of self-assembly that would indicate how to optimize the yield of complex targets, particularly in \emph{multifarious} systems, which are capable of assembling different target structures depending on the experimental conditions. Many of these theoretical studies investigate self-assembly within the framework of equilibrium statistical mechanics, in which the thermodynamic stability of a desired target state is maximized. 
Successful self-assembly has been achieved in various model systems by optimizing, according to this criterion, the subunit concentrations \cite{Murugan2015}, interaction strengths \cite{Hormoz2011, Zeravcic2017},  particle shape \cite{Damasceno2012, Sacanna2010}, and bond specificity \cite{Wang2012}.

Despite the successes of this approach, it suffers from a fundamental limitation: the thermodynamic stability of a structure does not guarantee its kinetic accessibility in the finite timescales available to experiments, due to the presence of long-lived intermediates along self-assembly pathways \cite{Hagan2006, Wilber2007, Grant2011, Palma2012, Whitelam2015, Bisker2018}. Finite-time assembly yields depend on a competition between thermodynamic and kinetic effects --- thermodynamic stability of a target structure and rapid nucleation are favored by strong interactions, whereas correction of misassembled subunits requires weak interactions. Identifying the optimal trade-off between these requirements can be difficult, both computationally and experimentally \cite{Trubiano2021}. 

As a result, a different route to achieve self-assembly has been proposed: driving the system out of equilibrium \cite{Das2021, Das2022} by using a time-varying protocol for system parameters \cite{Nguyen2016, Sherman2016, Heinen2015, Taylor2012, Tagliazucchi2014}. For example, temperature protocols with heating and cooling cycles have been shown to enhance control over crystal size distribution in crystallization experiments \cite{Snyder2007, Bakar2009} as well as selectively assemble a structure distinct from the global free-energy minimum in lattice simulations of a multi-component mixture \cite{Bupathy2022}. 
There is a broad range of experimental systems for which this approach can and has been applied; strand displacement reactions can tune interaction strengths between DNA-coated colloids \cite{Rogers2015, Gehrels2018}, light-activated interactions can be tuned via spatiotemporal intensity protocols \cite{Stenhammar2016}, and system properties such as temperature, pressure, and concentrations can be controlled within microfluidic devices \cite{Dou2017, Chang2018, Zhang2021}. 
While the technology to implement such protocols is available, the field lacks an efficient framework to rationally design and optimize them. `On-the-fly' methods \cite{Lindquist2016}, such as the statistical physics design engine \cite{Miskin2016}, and machine learning approaches, such as automatic differentiation of molecular dynamics trajectories \cite{Engel2022}, have been successful for select systems, but the high computational cost of sampling a large parameter space under experimentally relevant conditions is prohibitive for most systems. 
To bridge this gap, we seek to develop a method to efficiently compute optimal time-dependent protocols to maximize the yield of a chosen target state, even for challenging systems such as those exhibiting long assembly timescales, kinetic trapping, or competing metastable states.

Our approach relies on the construction of Markov State Models (MSMs) \cite{Husic2018, Perkett2014}. MSMs are a powerful tool for coarse-graining the dynamics of complex systems into a reduced-order form that is tractable to analysis and allows characterizing the system dynamics on timescales that are orders of magnitude longer than those accessible to straightforward simulations \cite{PANDE2010, Schwantes2014, Hummer2015, Husic2018, Suarez2021}. 
We show that, by taking advantage of properties satisfied by general MSMs \cite{Norris1997, Gardiner2004}, we can analytically and efficiently evaluate derivatives of state probabilities with respect to tunable system parameters for use in gradient-based optimization. Although our approach has features in common with previously developed methods to compute feedback control policies for self-assembly with real-time system sensors (e.g. \cite{Juarez2012, Tang2016, TANG2017, GROVER2019}), it avoids the potentially expensive dynamic programming calculations in those methods by directly considering the final time target state probability in the objective function.  
We first demonstrate the method on a semi-analytic Markov model for the assembly of a short polymer of six colloidal particles in two dimensions. For this system, we can optimize the time-dependent interactions between different particle types to achieve selective, high yield assembly of multiple structures, including the stable equilibrium structure, metastable rigid structures, and floppy structures that are only transient for most constant parameter sets. 
Our results pose an interesting choice of tradeoffs for experimental systems; we show that tuning a time-dependent protocol and increasing subunit complexity (i.e. the number of distinct subunit types) can have similar results on yield, but one of these choices may be easier to implement experimentally. 

To investigate generality, we also demonstrate our algorithm on a very different system --- a two-parameter model for capsid assembly on a spherical nanoparticle. 
For this example, there is no analytical form for the transition matrix. We show how simulations and interpolation can be used to make such complex problems suitable for our optimization algorithm.
By combining our algorithm with the use of radial basis function interpolation within parameter space, we develop a framework to dramatically reduce the computational requirements for protocol optimization. The method is also robust and reusable in comparison to existing methods; simulations do not need to be rerun to change the initial or target states in the optimization, as would be the case for `on-the-fly' methods. 
Using this procedure, we compute time-dependent protocols to assemble a target state that is highly kinetically inaccessible; there is approximately a $60\%$ gap between the estimated equilibrium and dynamical yields. For timescales we are able to simulate by brute force dynamics, our computed time-dependent protocols increase the target yield by greater than twofold over the best constant protocol, in the same amount of assembly time. The MSMs also allow for probing of longer timescales, for which we predict optimal time-dependent protocols that achieve yields within $1\%$ of equilibrium, in orders of magnitude less time than required to approach equilibrium under a constant protocol. 

Importantly, the method is generally applicable to any system that can be approximated by an MSM and optimized by an objective function involving state probabilities, including multifarious \cite{Murugan2014, Zwicker2022, Ben-Ari2021, Osat2022, Jacobs2021, Mohapatra2016} and reconfigurable assembly systems \cite{Das2022, Kohlstedt2013, Ortiz2014, Phillips2014, Young2012, Nguyen2010, Nguyen2018, Mann2009, Solomon2010, Long2014}. Moreover, it can be used as a highly efficient parameter estimation tool for both constant and time-dependent protocols.

\begin{figure*}[h!tb]
\centering  
\includegraphics[width=0.95\linewidth]{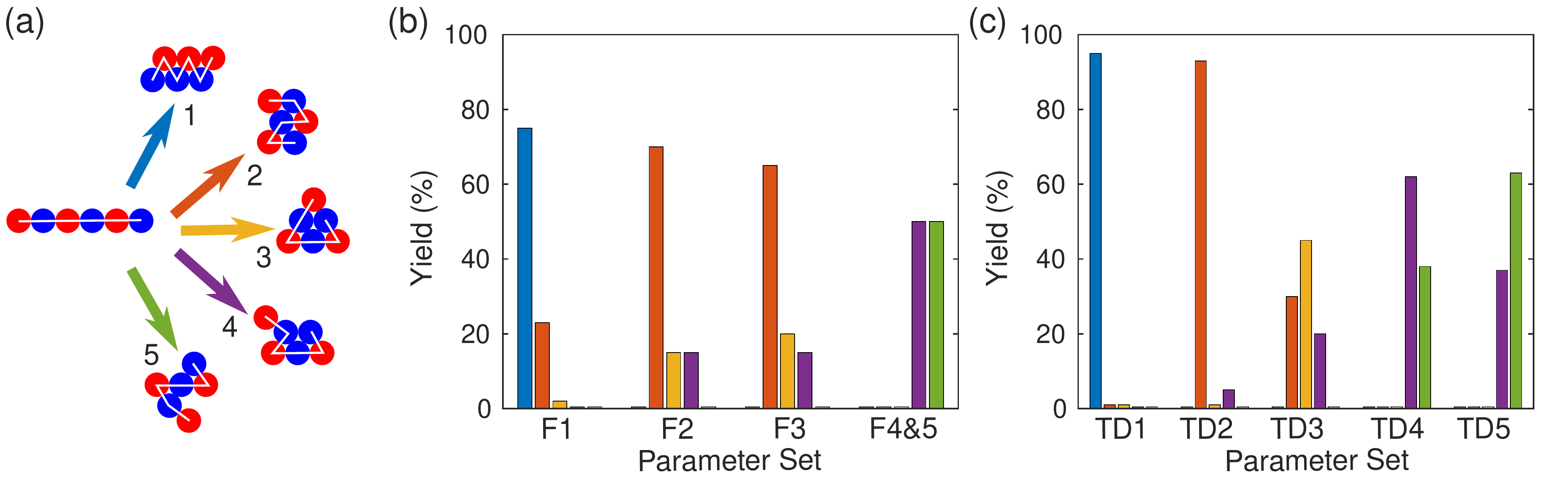}
\caption{ Visualization and statistics for the colloidal polymer system. 
(a) Depiction of the initial colloidal chain with alternating red-blue particle types and fixed backbone (shown in white), as well as example configurations of five states for which we optimize yields. Note that only a single representative permutation is shown for each state, but reported yields are combined over all consistent permutations. 
(b) Histograms of the yield for each of the five states in (a) (color-coded) for four sets of fixed parameter values. The number in the parameter set label indicates which state's yield is being maximized. Yields are computed by averaging over $600$ Brownian Dynamics trajectories with the given parameter values. The final time is $\TF = 5$ for state $1$, $\TF = 7.5$ for state $3$, and $\TF = 10$ for the others, reported as a non-dimensional simulation time. 
(c) Same as (b), but the parameter sets are now the optimal time-dependent protocols. 
} 
\label{fig:colloidModel}
\end{figure*}

\section{Methods}
\subsection{Theoretical Setup}
We begin by assuming the process of interest is described by a Markov State Model over the discrete state space, $\mathcal{S}$.
 For a given lag time, $\tau$, we define a temporal discretization with $t_n = n\tau$ for $n = 0, \dots, N$. The time-dependent protocol will be piece-wise constant, represented as $\{\theta_n\}_{n=0}^{N-1}$, where $\theta_n$ is a constant parameter value from time $t_n$ to $t_{n+1}$. The transition matrices are then given by $\mat{P}^n = \mat{P}(\theta_n)$, which propagate the state probabilities from $t_n$ to $t_{n+1}$. Extension to multiple control parameters, $\vec{\theta}_n$, as we use in the second example below, is straightforward.

Let the row vector $\niceVec{p}{n}$ denote the probability distribution over the system states at time $t_n$. The evolution of this probability is governed by the forward Kolmogorov equation, 
\begin{equation} \label{eq::fke}
    \niceVec{p}{n+1} = \niceVec{p}{n} \mat{P}(\theta_n), \qquad \niceVec{p}{0} = \vec{p}_0.
\end{equation}
For a set of target states, $B \in \mathcal{S}$, we are interested in optimizing $P_B^N = \sum_{i \in B} p_i^N$, subject to the constraint that the protocol does not change too rapidly. To this end, we maximize the objective function 
\begin{equation} \label{eq::obj_fun}
    \Phi[\theta] = P_B^N - \frac{\lambda}{2} \sum_{n=0}^{N-1} \left(\frac{\theta_{n+1} - \theta_n}{\tau}\right) ^2,
\end{equation}
where the second term is a smoothing penalty function whose strength is controlled by $\lambda > 0$. We will report this smoothing parameter as $\lambda^* = \lambda / (N\tau^2)$, which is normalized with respect to the timescales of the system. 
It is straightforward to add additional terms to Eq.~\eqref{eq::obj_fun}, for example to limit the magnitude of the control parameter or maximize the speed of assembly.

\subsection{Protocol Optimization}
We compute the gradient of $P_B^N$ via an adjoint method \cite{Plessix2006, Giles2000}. The adjoint equation is the backward Kolmogorov equation, 
\begin{equation} \label{eq::bke}
    \niceVec{F}{n} = \mat{P}(\theta_n) \niceVec{F}{n+1}, \qquad \niceVec{F}{N} = \vec{1}_B,
\end{equation}
where $F^n$ is the adjoint variable at time $t_n$, and $\vec{1}_B$ is the indicator vector for the set $B$. This equation is prescribed at a final condition and solved backward in time. By solving Equations \eqref{eq::fke} and \eqref{eq::bke} for all time, the gradient components can be computed (see the SM for details) as
\begin{equation} \label{eq::gradient}
    \frac{\partial P_B^N}{\partial \theta_k} = \niceVec{p}{k} \frac{\partial \mat{P}^k}{\partial \theta_k} \niceVec{F}{k+1}. 
\end{equation}

The derivative of the penalty term with respect to $\theta_k$ can be interpreted as a discretization of the second time derivative of the protocol. The gradient descent update step can then be seen as solving a discretized diffusion equation, where the gradient of $P_B^N$ acts as a source term. For stability reasons, we solve this equation with the IMEX \cite{Ascher1995} scheme
\begin{equation} \label{eq:gradient_descent_LS}
\theta_{k}^{j+1} = \theta_k^{j} + h \frac{\partial P_B^{N,j}}{\partial \theta_k}  + \frac{\lambda h}{\tau^2} \left(\theta_{k+1}^{j+1} - 2\theta_k^{j+1} + \theta_{k-1}^{j+1}\right), 
\end{equation}
where $j$ indexes the iteration number, $P_B^{N,j} = P_B^N[\theta^j]$, and $h$ is a step-size to be chosen. See the SM for details.

\subsection{Model Systems}


We demonstrate our optimization algorithm on two systems, which involve different levels of complexity to construct the MSM. The first example is a simple model for short polymer chains constructed from colloidal particles with programmable interactions, which enables analytical construction of an MSM. The second example is a model for viral capsid assembly in which two types of conical subunits assemble on the surface of a spherical nanoparticle. In this case we construct an MSM by estimating transition rate matrix elements from an ensemble of short unbiased simulations.

\subsubsection{Colloidal Chain Folding}

The first model describes the folding of an initially linear polymer made up of six colloidal particles in two dimensions. The folding problem has many features in common with assembly, including analogous thermodynamic and kinetic effects,
and provides a simple illustration of the optimal control method.
The interactions between each particle in the chain can be programmed, for example by coating the surface of the particles with strands of DNA \cite{Wang2015}. Further, these interactions can be varied in time by modulating the melting temperatures of different strand types \cite{Garima2010} and applying a temperature protocol.  
Depending on the choice of interaction strengths, the system can fold into one of three ground states as well as a number of floppy states, which have been characterized both experimentally \cite{McMullen2018, McMullen2022} and via theory and simulation \cite{Trubiano2021}. 
It has been shown for this system, as well as many others \cite{Hormoz2011, Russo2022, Jacobs2016, Zeravcic2014}, that increasing the number of subunit types allows for more efficient assembly of a target structure. However, increasing the number of subunit types generally makes systems more susceptible to kinetic traps and thus more sensitive to parameter variations. Further, additional species require greater costs in materials development and synthesis. 
Here we aim to investigate how time-dependent interaction strengths can boost folding yields, as well as identify tradeoffs between subunit complexity (number of different subunit types) and protocol complexity.

To study the minimal subunit complexity case that allows for multiple free energy minima, we allow for only two types of particles, which we will distinguish as red and blue. We assume the initial chain has an alternating ordering of these particles, i.e. red, blue, red, blue, and so on. 
Figure \ref{fig:colloidModel}(a) shows this initial chain as well as five potential folded structures. Structures $1$ through $3$ are the rigid ground states for this system, while structures $4$ and $5$ are floppy structures that are typically folding intermediates, but can be stabilized by setting some of the interactions strengths to zero. 
Figure \ref{fig:colloidModel}(b) shows histograms of the yields of each of these structures under a number of parameter sets, which consist of fixed values for each of the three interaction strengths; $\ERR$, $\ERB$, and $\EBB$, which are the well-depths for a short-ranged Morse potential. We denote parameter set $i$ as the one that maximizes the finite-time yield of structure $i$. We find that relatively large yields are already possible for most structures using fixed protocols, except for structure $3$, the triangle. While it is possible to further boost yields of some of these structures by changing the particle type distribution, we investigate what can be achieved using this fixed ordering along with time-dependent protocols. 

The model we use to construct an MSM for this system was previously developed to study equilibrium folding of colloidal chains \cite{Trubiano2021}. 
To summarize the model, a state space is defined by enumerating all possible adjacency matrices describing bonds between the six particles. For each adjacency matrix, the rate of forming a new bond and the probability distribution for which bond forms first are estimated using Brownian Dynamics simulations. This specifies all the forward rates of a rate matrix, which are independent of interaction strength. 
For a given set of interaction strengths, the equilibrium probabilities of each adjacency matrix are measured using a Monte Carlo sampler. The backward rates are then set from the forward rates by imposing detailed balance with respect to the estimated equilibrium measure. A re-weighting procedure can then be used to evaluate the backward rates for other values of the interaction strengths. 
The resulting rate matrix can be converted into a probability transition matrix by exponentiation, at which point the optimization can be performed as described above.

These simulations are performed in non-dimensionalized form. Positions, energies, and time are respectively scaled by particle diameter $\sigma$, $\kt$, and  a reference time $\tc$. The resulting equations have a non-dimensional diffusion coefficient $\epsilon = D \tc/\sigma^2$, where $D$ is the dimensional diffusion coefficient. We set $\epsilon=1$, and use experimentally measured values $\sigma=1.3\mu$m and $D=0.065\mu \text{m}^2/$s for DNA coated colloids \cite{Perry2015} to obtain an order-of-magnitude estimate of $\tc\approx 17$s. 

\begin{figure*}[h!tb] 
\centering  
\includegraphics[width=0.95\linewidth]{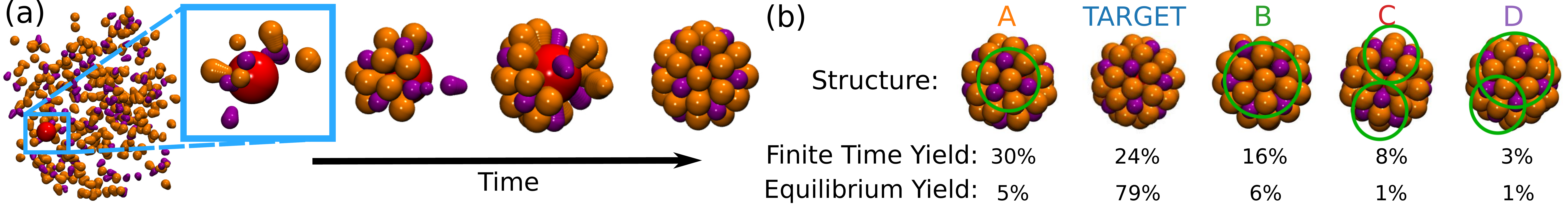}
\caption{Visualization and assembly statistics for the cones system. 
(a) Snapshots along a typical trajectory. The first image shows the entire simulation domain, consisting of pentamer subunits (purple), hexamer subunits (orange), and a spherical nanoparticle (red). The following images show a zoomed-in view of the nanoparticle as assembly progresses. 
(b) Visual renderings of the most common end states for simulations with fixed interactions set to $\EHP = 1.35 \kt$ and $\EHH = 1.475 \kt$, along with their finite-time yields at $\TF=8\times 10^5 \tzero$, estimated from $150$ independent simulations. The target capsid is the second most common end state.
The other states differ from the target due to the presence of one or more defects, which are circled in green. 
The bottom row on the right is the equilibrium yield for each structure, estimated by constructing an MSM and computing the stationary distribution. 
} 
\label{fig:snapshots}
\end{figure*}

\subsubsection{Conical Subunit Assembly On a Nanoparticle}

We study a system adapted from Ref. \cite{Lazaro2018}, consisting of two types of conical subunits that assemble into capsids on the surface of a spherical nanoparticle. 
The subunits are rigid bodies comprised of six beads of increasing radius, consistent with a cone angle, $\alpha$. Each of the four interior beads interacts attractively with the corresponding bead on other subunits, through a Morse potential. The innermost bead has an attractive Morse interaction with the nanoparticle. 

The two subunit types have relative diameters such that they correspond to pentamers (P) and hexamers (H) in the lowest energy icosahedral capsid structure \cite{Caspar1962}. 
There are two types of interactions between the subunits; an H-H attraction and an H-P attraction. The strength of these interactions can be tuned by varying the well-depth of the corresponding Morse potential, $\EHH$ and $\EHP$, respectively. Interaction strengths are chosen in the range $1.2 \kt$ to $1.8 \kt$, which can facilitate assembly on the nanoparticle surface, but typically is too weak to drive nucleation in the bulk.
The strength of the attraction to the nanoparticle is held constant at 
$\ENH=7\kt$ for hexamers and $\ENP=6.3\kt$ for pentamers.  The system contains one nanoparticle, $86$ pentamer subunits and $214$ hexamer subunits, so that subunits are in excess and the chemical potential of free subunits remains nearly constant throughout the assembly process. We consider a cubic box with side length $120 \lzero$, where $\lzero=1$nm is set as the unit length scale for the system. 
Assembly simulations are performed using the Brownian dynamics algorithm in HOOMD \cite{Anderson2020} with periodic boundary conditions. Times are measured in terms of the HOOMD derived unit, $\tzero = \sqrt{m/\kt} \lzero$, where $m$ is the subunit mass.  We also construct an MSM for the system by estimating the transition matrix elements from these Brownian dynamics simulations (section~\ref{sect:MSM_main}). Further details on the simulations are provided in the SM. 

The structures with the lowest potential energy (per particle) for this system are capsids with either $T=4$ icosahedral symmetry or $D5$ symmetry, which both contain $30$ hexamer and $12$ pentamer subunits with the same number of H-H and H-P bonds \footnote{ Measuring the average energy of assembled capsids in simulations indicates that the $D5$ structures are slightly ($< 0.1 \%$) lower total energy than $T=4$ capsids. }.
However, in finite-time dynamical simulations assembly usually results in asymmetric structures that, although containing about 42 subunits, have one or more defects.  
Figure \ref{fig:snapshots} depicts the five most probable end structures after $\TF = 8\times10^5\tzero$, for the parameters that maximized the yield of symmetric capsids ($T=4$ and $D5$). The observation time $\TF$ was chosen as the timescale after which the yield only grows logarithmically for most parameter sets that we focus on. 
We also compare the finite-time yields to the equilibrium yields, which we estimated using the MSM. This comparison shows that the system is far from reaching equilibrium at $\TF$, and from the MSM we estimate that approaching equilibrium would require a timescale of approximately $60\TF$. 

These observations are consistent with the  competition among thermodynamic and kinetic effects that arise for constant assembly driving forces. For very low subunit-subunit interactions or subunit concentrations, assembly is either thermodynamically unfavorable or does not nucleate on relevant timescales, while overly high interactions or concentrations  cause the system to become trapped in metastable structures. Since subunit-subunit interactions must be broken to exit a metastable state, reconfiguration timescales increase exponentially with interaction strengths. In our system, while the $T=4$ and $D5$ structures are energetically favored, the other metastable states that we observe are more kinetically accessible under interaction strengths that enable nucleation on relevant timescales and thus occur in most assembly trajectories.

These competing effects suggest that assembly yields and fidelity could be  enhanced by time-dependent sequences of interactions, which can drive rapid nucleation while also avoiding kinetic traps by facilitating reconfiguration from metastable states into the ground state(s). To investigate this possibility, we apply our optimization algorithm to the cones system. Our objective is to compute an optimal time-dependent protocol for $\EHH$ and $\EHP$ interactions that maximizes the yield of the symmetric $T=4$ and $D5$ capsids at a specified observation time.

\begin{figure}[t]
\centering  
\includegraphics[width=0.95\linewidth]{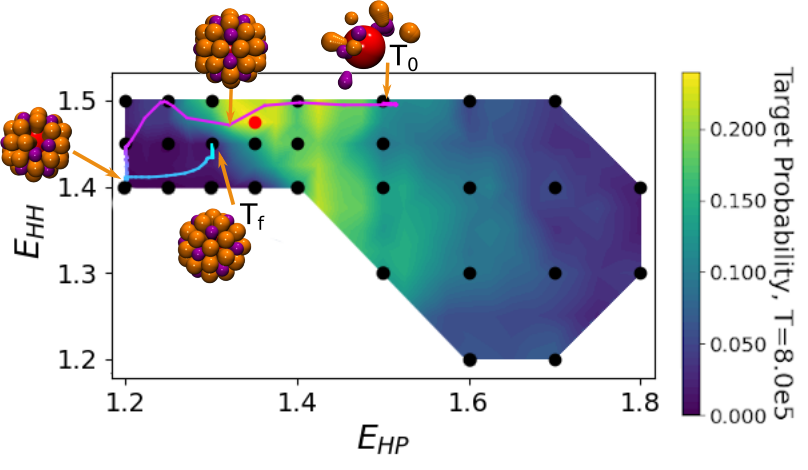}
\caption{Feasibility set for the cones system. The allowed parameter values for the optimization are shown within the bounded region of parameter space. Black nodes represent parameter values for which we constructed local MSMs. Global transition matrices are interpolated for the optimization and to evaluate and plot the assembly populations as a function of time. 
The red point indicates the parameters that achieve the maximum target yield with fixed interactions. The optimal time-dependent protocol is traced by the curve, from purple ($T_0$) to light blue ($\TF$), with representative snapshots of the structures at various time points. 
} 
\label{fig:feasible_set}
\end{figure}

\subsection{Constructing Markov State Models} \label{sect:MSM_main}
To apply the protocol optimization algorithm in Eq. \eqref{eq:gradient_descent_LS}, transition matrices need to be evaluated for the possible protocol values that are encountered during maximization of the objective function. While this analysis is computationally efficient for the colloidal polymer system because we have a semi-analytic Markov model, we are not so fortunate for the cones model. Similarly, for most relevant systems it will not be possible to analytically evaluate the transition matrix. It would be computationally intractable to perform the sampling from unbiased simulations required to estimate the transition matrix at every candidate parameter value encountered during every candidate protocol sequence.  Therefore, we describe here a computationally efficient procedure for estimating an MSM over parameter space to apply our optimization algorithm. This process has three steps: (1) We  construct and validate \emph{local MSMs} for fixed values of the system parameters, distributed over the region in parameter space that assembly protocols are likely to visit (which we denote as the \emph{feasibility set}). (2) We perform an interpolation to construct a global MSM that can be evaluated for any intermediate values of the feasibility set. (3) During the optimization, if the protocol attempts to leave the feasibility set, we perform additional sampling from unbiased trajectories to expand the feasibility set as required. Similarly, if analysis indicates that interpolation errors are large within a subregion of the feasibility set, we perform sampling at additional parameter sets within that subregion.

To construct local MSMs for the cones system, we require a state space discretization; i.e., a mapping between simulation configurations and MSM microstates. We use the state space discretization $(\np, \nh, \IH)$, where $\np$ and $\nh$ are the number of pentamers and hexamers attached to the nanoparticle, and $\IH$ is equal to the number of hexamers in contact with precisely two pentamers. The target state is specified by the triplet $(12,30,30)$, which corresponds to a $T=4$ or $D5$ capsid. 
 We find that this three-variable description is insufficient for a subset of configurations, for which we augment the discretization with the variables $\bhh$ and $\bhp$, the number of hexamer-hexamer and hexamer-pentamer bonds, respectively. See the SM for details. 

We construct MSMs from unbiased dynamics trajectories by computing a row-normalized count matrix of the number of transitions between discrete states after a lag time, $\tau$. 
We determine that $\tau= 3125\tzero$, is sufficient to build converged local MSMs and capture the assembly dynamics (see the SM for detailed convergence and validation). 
We identify a region in parameter space that leads to productive assembly, and perform simulations at parameter values on an unstructured grid within that region to construct a collection of local MSMs. 
Figure \ref{fig:feasible_set} shows this region for our example and the sampled parameters within it as black nodes. To avoid extrapolation errors, we restrict optimization to the feasible set. However, as noted above, the feasible set can be expanded by additional sampling if the protocol attempts to leave the region.

To construct an MSM during the protocol optimization at parameter sets between sampled nodes, we perform interpolation of the transition matrix entries using radial basis functions \cite{Fasshauer2007, Hines2016}. 
Since each parameter set may sample a disjoint collection of states, we first define a global discretized state space as the union of all states observed in each local MSM. 
For each pair of states with at least one transition among all parameter sets, we construct a list of their transition probabilities for each parameter set. We assign a probability of $0$ if the transition is not observed. 
Using a Gaussian kernel, we construct an interpolant for each non-zero transition matrix entry, which can be evaluated for any parameter values in the allowed region, and used to construct a global transition matrix. See the SM for details on constructing and evaluating these interpolants. Note that this process can be made more efficient by using information from sampling at all parameter sets for evaluation of a given transition matrix, for example by using transition-based reweighting analysis methods (e.g. DHAM \cite{Rosta2015}, xTRAM \cite{Mey2014} and dTRAM \cite{Wu2014}). 
For simplicity, we have not employed these protocols in this work.

\section{Results}

\subsection{Colloidal Chain Assembly}

\begin{figure*}[t]
\centering  
\includegraphics[width=0.99\linewidth]{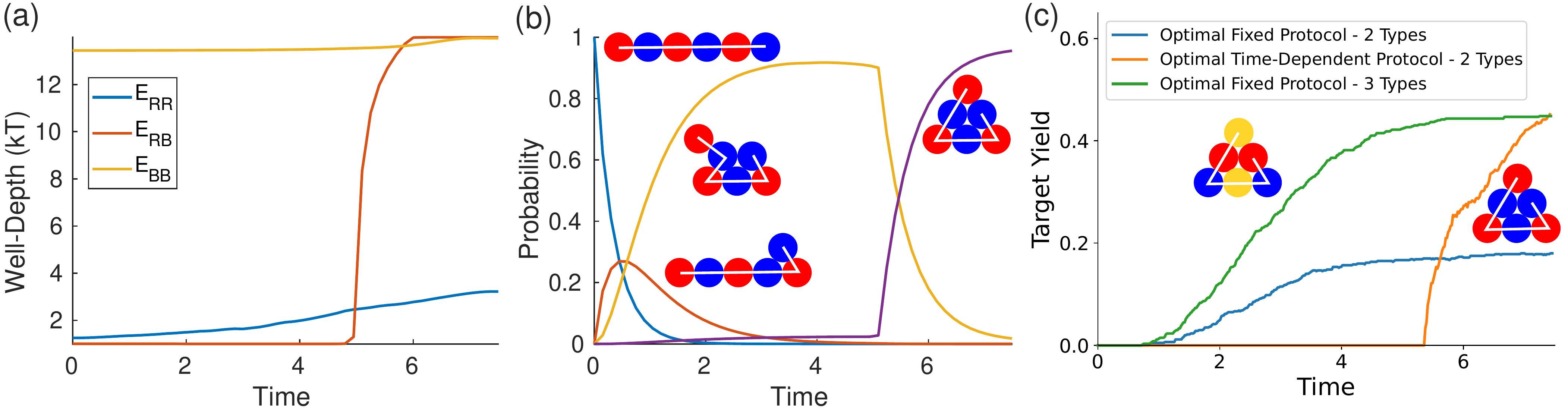}
\caption{Protocol optimization results and verification for the triangle state of the colloidal polymer. 
(a) Optimal time-dependent protocol for the three interaction parameters found by our algorithm. The initial guess was the optimal constant values for the like interactions, $\ERR=0.01\,\kt$ and $\EBB=14\,\kt$, and a linear protocol from $1\,\kt$ to $12\,\kt$ for $\ERB$.  
(b) State populations as a function of time for the optimal protocol, evaluated using the MSM. Only states with a maximum yield of over $0.2$ are shown.
(c) Time-dependent yield of the triangle state computed from Brownian dynamics. For the alternating chain, we compare the optimal fixed protocol yields with the optimal time-dependent protocol yields. We also show the yield for the optimal fixed protocol with \emph{three} particle types, and the corresponding optimal particle. Results are averaged over $600$ independent Brownian dynamics trajectories.
} 
\label{fig:triangleOpt}
\end{figure*}

We apply our optimization code to compute the time-dependent protocol that maximizes the yield for each structure shown in Figure \ref{fig:colloidModel}(a). 
We use different final times $\TF$ for each structure, choosing $\TF$ to be long enough for each phase of a protocol to approximately reach a steady state.  
We use a regularization parameter $\lambda^*=2*10^{-3}$ for structures $1${-}$3$ and $\lambda=2*10^{-4}$ for structures $4$ and $5$. We believe that it was it necessary to reduce $\lambda$ for states 4 and 5 because these are floppy structures that are difficult to stabilize.
In all cases, we use a line search with starting step size $h=1$, and set the lag time to $\tau=\TF/50$. 
The resulting yields are shown in Figure \ref{fig:colloidModel}(c), where the label number denotes which structure was being optimized for. Comparing these values to the optimal yields for fixed protocols in Figure \ref{fig:colloidModel}(b), we see improvements for each structure. We find that structures $1$ and $2$ can be formed with relatively high selectivity among the tracked structures, whereas the protocols for the other targets result in a spread of structures, but with the target being the maximal probability structure in each case. 

The triangle state, structure $3$, sees the largest gain in yield with the time-dependent protocol among the five. The yield more than doubles from around $20\%$ to just under $50\%$. 
The protocol is relatively simple, shown in Figure \ref{fig:triangleOpt}(a). It begins with a large $\EBB$ value, which we know to form structure $4$ with roughly $50\%$ yield. Then by switching on a large value for $\ERB$, the triangle state becomes the dominant endpoint for any trajectory that enters the structure $4$ state. 
Figure \ref{fig:triangleOpt}(b) shows the probability for the highest yield structures using this protocol, evaluated using the MSM, which shows the pathway taken to form the target. 
Note that the MSM predicts a nearly perfect yield, but simulation results in a yield of around $50\%$. This is due to a two-fold degeneracy in the adjacency matrix representation of states along this transition pathway. Whether the outer particle rotates clockwise or counter-clockwise to form bonds, the adjacency matrix representation is the same, but the resulting configurations either form the triangle state or a kinetic trap, reducing the maximum yield by a factor of two (see the SM for further discussion). 


Finally, Figure \ref{fig:triangleOpt}(c) shows the yield of the triangle state as a function of time for three protocols, evaluated from unbiased Brownian dynamics trajectories. We see the optimal fixed protocol for the alternating chain quickly levels off at around $20\% $ yield. In contrast, for the optimal time-dependent protocol, probability first accumulates in structure $4$, then very quickly forms the triangle once the red-blue interaction is turned on, achieving around a $50\%$ yield. We also compare these curves to the best fixed protocol that can be achieved by adding additional particle types. The minimal complexity solution turns out to be three particle types, in the configuration shown in Figure \ref{fig:triangleOpt}(c), which is as efficient as having all particles being unique \cite{Trubiano2021}. We see that we can achieve comparable assembly efficiency by using a time-dependent protocol with two particle types as we can by optimizing constant protocols with three particle types. These kinds of tradeoffs are important to characterize, as in many systems it is more economical to vary experimental conditions in time than it is to design and synthesize new subunit types with specified interactions. 

The optimal protocols and the Brownian dynamics verification of their predicted yield for each of the other structures in Figure \ref{fig:colloidModel}(a) can be found in the SM. Our computed protocols and expected yields for the rigid structures are consistent with recent experimental findings \cite{McMullen2022}.

\subsection{Conical Subunit Capsid Assembly}

First, we get a sense of how the assembly proceeds under fixed parameter values, with an observation time $\TF = 8\times10^5\tzero$, or $256\tau$. We construct the transition matrix interpolant  for parameter values inside the feasibility set (indicated by black points in Figure \ref{fig:feasible_set}), and then use it to compute target yields according to Eq.~\eqref{eq::fke}. The resulting yields are shown as a heat map in Figure \ref{fig:feasible_set}. 
We see that if the interactions are too weak, the target does not form due to thermodynamic unfavorability or slow nucleation. 
As the interactions strengthen, we reach the optimal assembly conditions ($\vec{E}=(\EHP, \EHH)=(1.35\kt, 1.475\kt)$, red point), 
where the target is stable, but subunits can rearrange to correct defects during assembly trajectories. Further strengthening the interactions reduces the target yield, as defect states can no longer rearrange on simulated timescales and the system becomes trapped in metastable states. This nonmonotonic dependence of yields on interaction strengths for fixed interactions is typical in self-assembling systems \cite{Hagan2006, Whitelam2015, Grant2011}.

\begin{figure}[t]
\centering  
\includegraphics[width=0.99\linewidth]{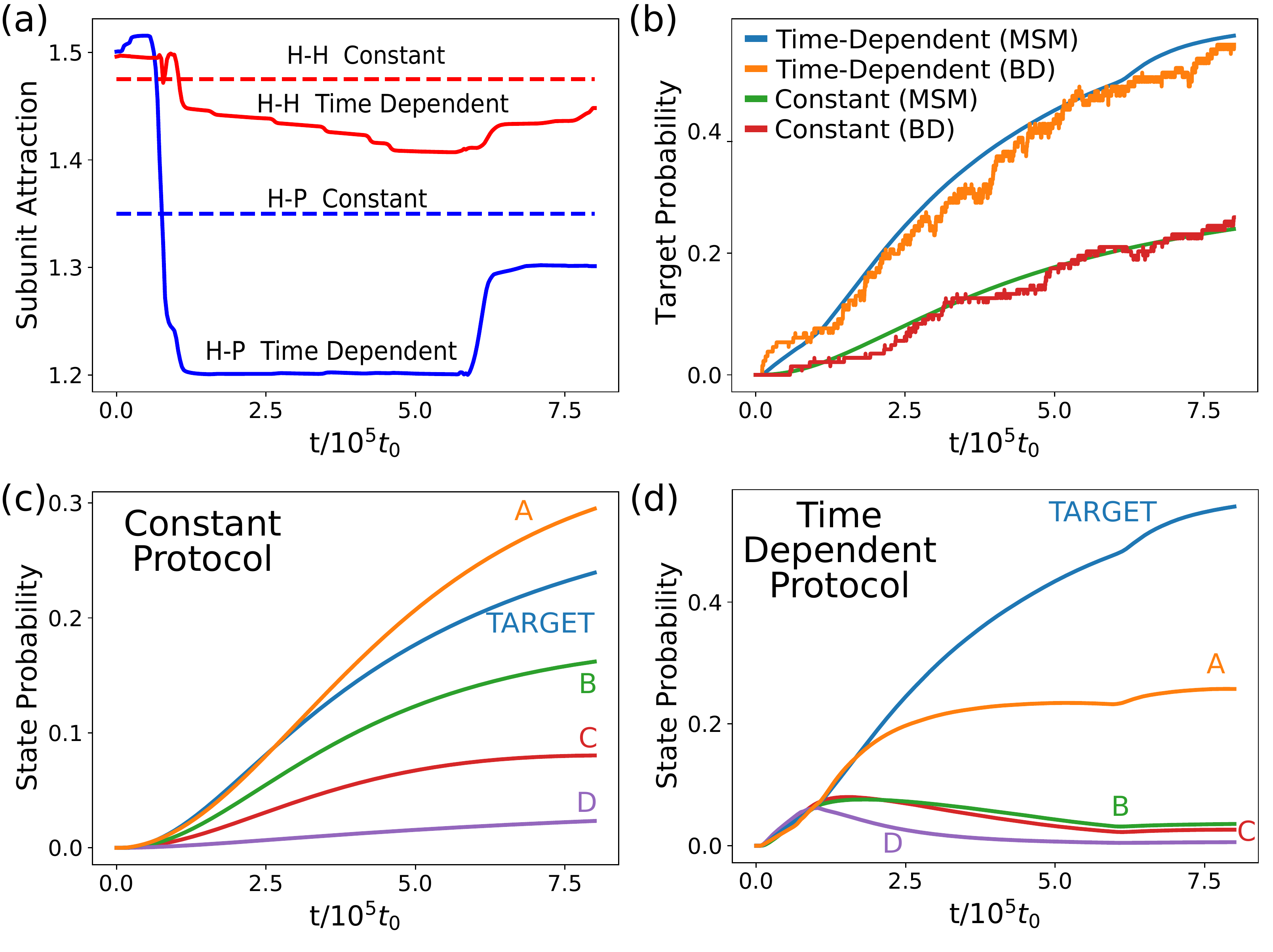}
\caption{Protocol optimization results and testing for the cones system. 
(a) The optimal time-dependent protocol found by our algorithm is shown as a solid line. 
The optimal constant protocol, determined in Figure \ref{fig:feasible_set}, is shown as a dashed line. 
(b) MSM estimates of the target probability as a function of time for the optimal time-dependent protocol and optimal constant protocol. The MSM estimates are verified by performing $150$ Brownian dynamics (BD) simulations performed with the optimal time-dependent or constant protocols respectively, and tracking the number of trajectories in the target state as a function of time. 
(c) MSM estimates of all common state probabilities using the optimal constant protocol. 
(d) MSM estimates of all common state probabilities using the optimal time-dependent protocol. 
}
\label{fig:prot_probs}
\end{figure}

Next, we apply our optimization algorithm to the capsid assembly problem using $\lambda^* = 4*10^{-3}$ 
and $h$ determined each iteration by a back-tracking line search. Our initial guess is a piece-wise constant protocol, $\vec{E}=(1.5 \kt, 1.5\kt)$ for the first $25$ lag times, and $\vec{E}=(1.3\kt, 1.5\kt)$ for the remaining time. The protocol converges to the result shown in Figure \ref{fig:prot_probs}(a). 
The time-dependence of this protocol is traced out in the parameter space in Figure \ref{fig:feasible_set}, and representative snapshots of the system are shown along the pathway. The sequence goes from an empty nanoparticle, to defective structure C, through a variety of transient structures, and finally the target structure. 

By propagating the time evolution of state probabilities with the MSM, we obtain estimates for the target yield as a function of time for each protocol (Figure \ref{fig:prot_probs}(b)). The optimal time-dependent protocol achieves a yield of about $55 \%$ at our chosen final time $\TF=8\times10^5\tzero$, compared to the optimal constant protocol yield of about $22 \%$. Thus, the time-dependent protocol enhances yields by more than two-fold in the same assembly time. 
To test these estimates, we perform Brownian dynamics according to the same protocols, and compare the computed yield curves (averaged over 150 independent Brownian dynamics simulations) to the MSM prediction (Figure \ref{fig:prot_probs}(b)). The MSM estimates match the Brownian dynamics results well, even for the time-dependent protocol. This close agreement demonstrates that, with sufficient sampling for estimating transition matrices and sufficient coverage of the feasibility set, the radial basis function interpolation enables effective MSM predictions for non-sampled parameter sets.

\begin{figure}[t]
\centering  
\includegraphics[width=0.99\linewidth]{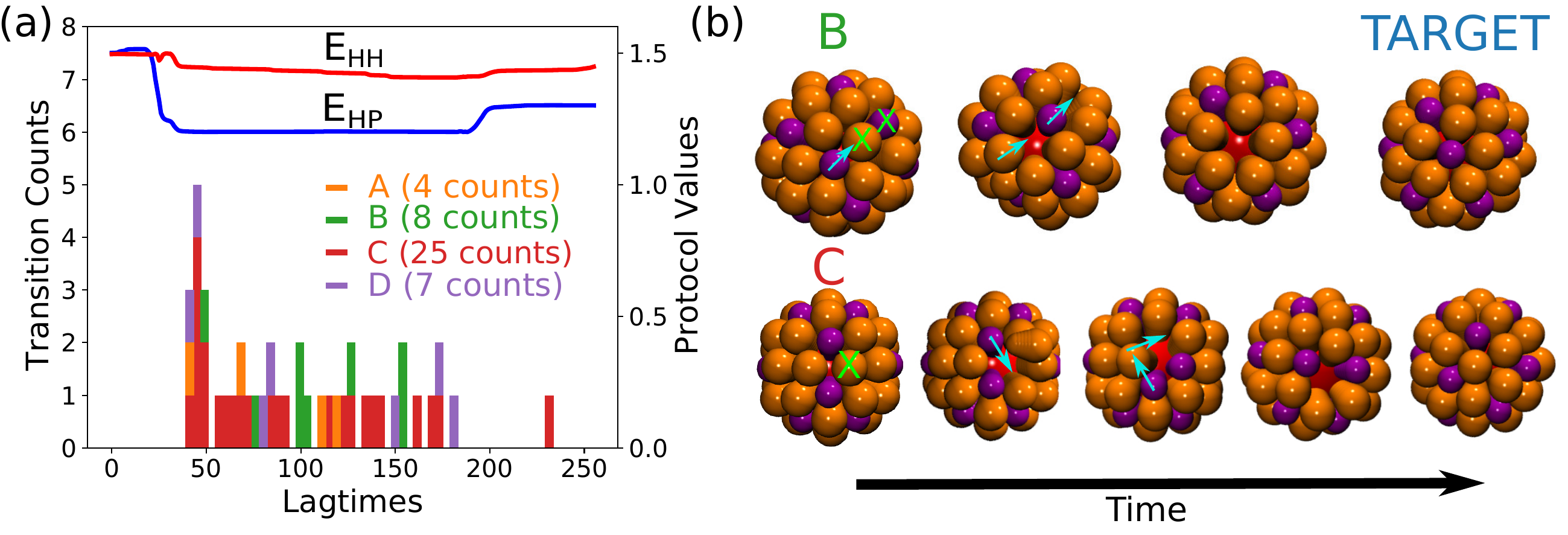}
\caption{ Analysis of how the optimal time-dependent protocol enhances yield. 
(a) Distribution of times when transitions from the competing states to the target state begin in trajectories with the optimal time-dependent protocol (from Figure \ref{fig:snapshots}(b)).
(b) Examples of transition pathways from states B and C into the target state, observed under the optimal time-dependent protocol. Green X's denote that a particle has dissociated in the next snapshot, and blue arrows denote the reconfiguration of associated subunits that results in the next snapshot. 
}
\label{fig:transitions}
\end{figure}

Figures \ref{fig:prot_probs}(c) and \ref{fig:prot_probs}(d) show yield curves for each of the high-population structures (those shown in Figure \ref{fig:snapshots}(b)) for the optimal constant and time-dependent protocols respectively. 
For the optimal constant case, we see the yield curves are all nearly flat near time zero, reflecting the length of time required for nucleation and growth of capsids. 
Further, the yield curves for all structures increase monotonically over the simulated timescale, implying that once one of these structures is reached, transitions to the others are unlikely. Consistent with this conclusion, only $4$ out of $150$ simulation trajectories exhibit a transition between labeled states, which all occur from state C to the target. 
In contrast, the time-dependent protocol enables rapid nucleation and growth of capsids. Each tracked state accumulates approximately $10\%$ yield, at about the same rate, during the initial growth phase. During the next phase, where interaction strengths are reduced, states B, C, and D are destabilized and begin to decrease in probability while state A continues to accumulate, but more slowly. The target state, however, quickly increases in yield, indicating that transitions are occurring from these destabilized structures to the target structure. In simulations, we observe $44$ of $150$ trajectories with such transitions, occurring from all of the A, B, C, and D states into the target. 
The distribution of times along the optimal protocol when these transitions begin is plotted in Figure \ref{fig:transitions}(a), which shows that the transitions are concentrated when the interactions first become weak, and that state C is the initial state for most transitions. 
Figure \ref{fig:transitions}(b) shows two example 
transition pathways. Both pathways begin with a subunit detaching from the nanoparticle, allowing rearrangement of the nearby subunits. While hexamers are involved in the rearrangements, pentamer rearrangements seem to drive the transitions, which explains why the optimal protocol decreases $\EHP$ more than $\EHH$. 

The rapid transition between weak and strong interactions in the optimal time-dependent protocol is qualitatively consistent with Fullerton and Jack's results on colloidal cluster growth \cite{Fullerton2016}, which considered simpler protocols defined by two interaction strengths and a single time to switch between them. To investigate how much additional enhancement a full time-dependence achieves in comparison to a piece-wise constant protocol, we converted our protocol to a simple two-step protocol. 
We used the time of the large jump as the transition time, at which point the interactions switch from $\vec{E}=(1.5\kt,1.5\kt)$ to $\vec{E}=(1.2\kt,1.4\kt)$. This protocol gives a yield of about $49.9 \%$, so we estimate that the fully time-dependent protocol boosts the yield by an additional $5\%$. While this further improvement is relatively small, our efficient optimization algorithm provides a converged, fully time-dependent protocol much faster than it would take to sample all possible three parameter step-function protocols. In particular, one gradient descent iteration requires approximately the same computation time as sampling two step-function protocols, and we typically observe convergence in $20-40$ iterations (for a good initial guess protocol).

\begin{figure}[t]
\centering  
\includegraphics[width=0.99\linewidth]{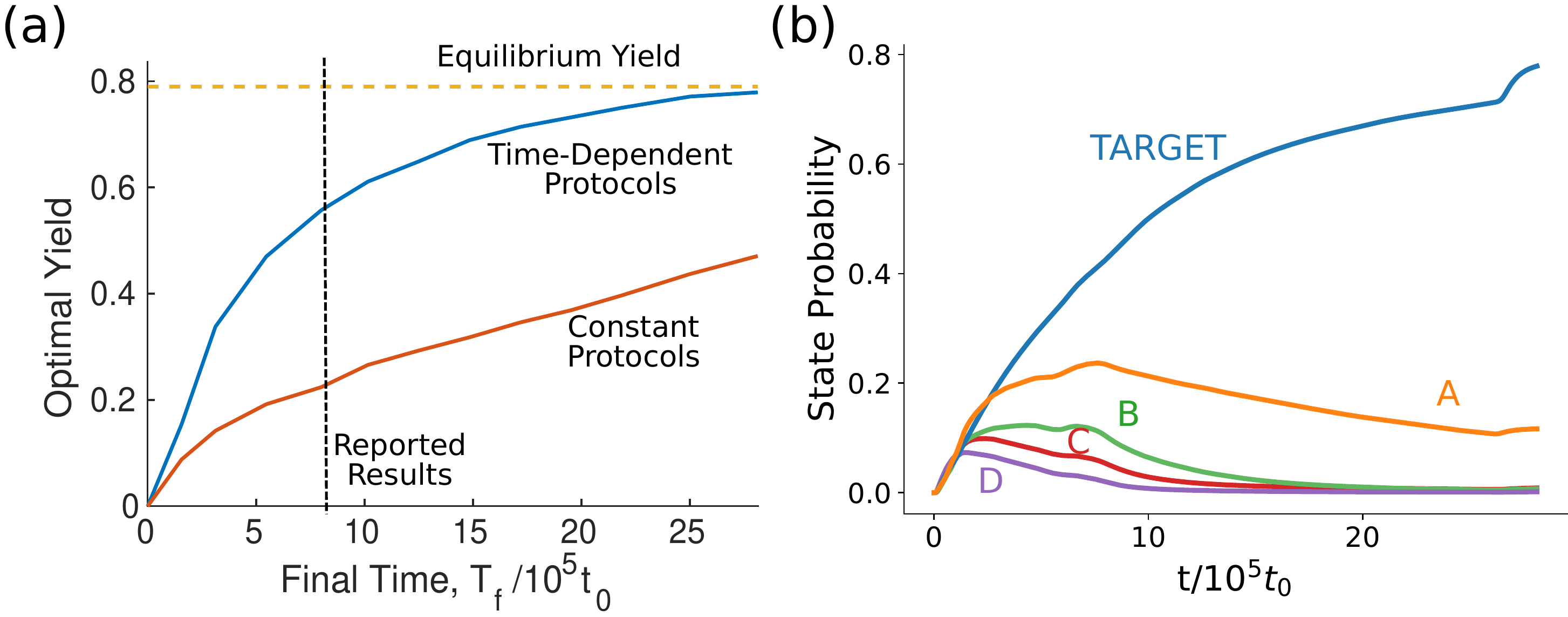}
\caption{ Effect of the final time on optimal yields.  
(a) The optimal target yield from our computed constant and time-dependent protocols as a function of the final time. For reference, the yellow dashed line shows the estimated equilibrium target yield as in Figure \ref{fig:snapshots}(b), and the black dashed line shows the final time used to generate our prior results. 
(b)  MSM estimates of all common state probabilities using the optimal protocol for $\TF = 900\tau \approx 28.12 \times 10^5 \tzero$. 
}
\label{fig:yield_vs_Tf}
\end{figure}

\textit{Effect of final time and regularization.} Thus far, we have performed the protocol optimization using a final time $\TF=8\times10^5\tzero$, since this is tractable for testing the predicted protocols with Brownian dynamics simulations. One of the benefits of constructing MSMs is that it enables the probing of timescales beyond what is feasible for simulation. We take advantage of this to compute optimal protocols for larger values of $\TF$. 
We find that the resulting protocols are all similar; they all spend about the same amount of time in the strong interaction phases at the beginning and end of the protocol, and the extra assembly time gets allocated to the weak interaction phase as $\TF$ increases. This is likely because we only optimize for the target yield at the final time, without enforcing constraints on assembly rate. 
Figure \ref{fig:yield_vs_Tf}(a) shows the optimal time-dependent and constant target yields as a function of the final time. We see that for constant protocols, the optimal yield increases approximately linearly, but slowly, while the time-dependent optimal yield increases quickly initially and then levels off. As it levels off, it seems to approach the estimated equilibrium yield for the target state, coming within $1\%$ of it when $\TF = 900\tau$.
To determine what sets the timescale for this approach to equilibrium, we plot the MSM estimated probability of each common state in Figure \ref{fig:yield_vs_Tf}(b). States B, C, and D reach steady state probabilities along this protocol, but state A is still decreasing to its equilibrium value near the final time. State A is seen to be the slowest to transition to the target, so this transition rate is what sets the necessary time-scale for the optimal time-dependent protocol to drive the system towards its equilibrium statistics. This suggests another mode for further optimization; search for new regions in parameter space that either minimize assembly of state A or allow for quicker transitions out of it. 

The optimal protocol and the resulting maximum yield depend on the regularization constant $\lambda$ in Equation \eqref{eq::obj_fun}. The results shown here used $\lambda^*=4*10^{-3}$; by decreasing $\lambda^*$ 
by a few order of magnitude, we observe maximum yields that increase by up to $2\%$, but also involve protocols which are much less smooth than the one shown in Figure~\ref{fig:prot_probs}(a). Thus, one can adjust $\lambda$ depending on the trade-off between difficulty in implementing abrupt changes in the control parameters and increase in yield. See the SM for an analysis of how $\lambda^*$ affects the optimal yield and protocol smoothness.

The optimization also converges to different results based on the initial guess, suggesting that we have not found the global maximum. We performed additional optimizations that were initialized with various piece-wise constant protocols, as well as a linear protocol between the optimal start and end values. The results are qualitatively similar in each case; interactions go from strong to weak to strong, but at different times, with target yields ranging from $45\%$ to $55\%$. These results obtained from different initial guesses suggest that our protocol is near-optimal, but that it would be beneficial to use alternative optimization algorithms better suited to finding global extrema on rugged landscapes.

\section{Conclusions}
Markov State Models (MSMs) have been widely adopted by the molecular dynamics community for their ability to probe long-time kinetics in a way that is human-interpretable. By leveraging these properties of MSMs, as well as properties of the underlying transition matrices, we have developed an optimization algorithm that can efficiently compute time-dependent parameter protocols that maximize the finite-time probability of observing a chosen target state. Furthermore, in the (typical) case when an analytic form for the transition matrix cannot be obtained, we have described a sampling and interpolation procedure that enables constructing MSMs as a continuous function of the system control parameters after performing sampling at only discrete parameter sets. The resulting MSM can then be used for optimization. 

We have tested the method and evaluated its performance on two systems. For a short polymer of colloidal particles, time-dependent interaction strengths between particle types can be tuned to drive selective folding into a number of different cluster geometries. Importantly, we show that the time-dependent protocols can be used to selectively enhance the yield of not only the equilibrium ground state, but also metastable states including floppy structures which are typically transient under constant protocols. For some structures, it is possible to reduce the number of distinct particle types needed to achieve high folding yields under constant parameters, by instead tuning a time-dependent protocol. Thus, this approach could simplify experimental realizations of folding and assembly, since time-dependent protocols may be easier to implement in some systems than synthesizing new subunit types with specific interactions. For example, in systems with interactions mediated by DNA hybridization, Rogers et al. \cite{Gehrels2018,Rogers2015} showed that DNA sequences can be designed such that multiple sets of particle-particle interactions can be simultaneously tuned by varying temperature. 

We also consider a more computationally challenging system, conical subunits assembling around a spherical nanoparticle as a model for viral capsid assembly. In this example, the target yield is suppressed because kinetic traps corresponding to defective structures prevent many trajectories from attaining the ground state configuration on relevant timescales. We demonstrate that by optimizing time-dependent subunit-subunit interaction strengths, the maximum yield can be more than doubled in comparison to that achieved with constant interactions. 
The optimal protocol begins with strong interactions that drive rapid nucleation, but into primarily defective structures, followed by selectively weakening interactions to enable rearrangement into the stable target state. 
Using unbiased simulations, we verify that the MSM estimates are accurate to within a few percent yield, which is notable considering the potentially compounding errors arising from the discretization of state space when constructing the MSM and the interpolation error associated due to estimating transition matrices at discrete parameter sets. Importantly, one should not extrapolate outside of the sampled domain (feasibility set) when using the interpolation procedure. As an example, in the cones system extrapolation outside of the feasibility set (based on the nearby transition matrix values) predicts nucleation in some parameter regions where unbiased simulations exhibit no nucleation, leading to large errors in MSM predictions. Instead, the feasibility set can be enlarged by additional sampling if the optimization algorithm identifies additional regions of parameter space as important.

While we developed this algorithm in the context of folding or self-assembly protocols, the adjoint-based optimization algorithm is generically applicable to MSMs and any objective function involving a state probability. For instance, the same framework can be used as a highly efficient parameter estimation tool, by determining the protocol (e.g. parameters such as temperature, interaction strengths, and concentrations) that most closely reproduces an experimental yield curve. Such estimation can be performed for time-dependent or constant protocols. 
Alternatively, the algorithm can be used for multifarious \cite{Murugan2014, Zwicker2022, Ben-Ari2021, Osat2022, Jacobs2021, Mohapatra2016} or reconfigurable \cite{Das2022, Kohlstedt2013, Ortiz2014, Phillips2014, Young2012, Nguyen2010, Nguyen2018, Mann2009, Solomon2010, Long2014} assembly systems. By changing the initial and target states in the optimization, one can optimize for assembly of a wide range of target states or to promote transitions between particular states.

\textit{Combining optimal control theory with closed-loop feedback control.} The method described in this article uses an MSM and a single optimal control theory computation to generate a protocol that is optimal at the ensemble level. On average, we have shown that the assembly will proceed according to the MSM predictions, but due to the stochastic nature of self-assembling systems there will inevitably be some trajectories that quickly reach the target, some that take longer, and some that get trapped. Moreover, stochastic noise and experimental error could diminish the performance of the predicted protocol. Both of these limitations could be improved by combining our method with closed-loop feedback control, which has been shown to make assembly more robust by allowing the protocol to respond to the current state of the system, rather than the average state of the system at that time (e.g. \cite{Tang2016, TANG2017, GROVER2019}). Our method can be applied to closed-loop control by monitoring the current state of a simulation or experiment at prescribed time intervals. At each time interval, the current system configuration is set as the initial state in the optimization, and a new optimal protocol is computed.

%

\textit{Limitations of the method and outlook.}
In this work, we have considered systems in which the subunit chemical potential is irrelevant (for the folding problem) or nearly constant in time (since the subunits are in excess in the cones system). We will describe in a subsequent publication how the method can be modified to account for time-varying chemical potentials, such as occurs due to depletion of subunits in systems involving homogeneous nucleation or where subunits are not in excess.

The current implementation of our method also suffers from a few limitations. 
The first is scaling to a larger number of control parameters. 
Sampling and interpolation costs scale exponentially in the number of parameters. 
Fortunately, there is a desire to be economical in terms of system design, so many systems of interest will have only a few independent parameters. In higher dimensional parameter spaces, a more targeted sampling approach will be necessary; for example, sampling could be performed on-the-fly in regions indicated as important by the optimization algorithm.

State space discretization is another important consideration, as it is for constructing an MSM in any system. A good state space discretization must describe all relevant slow degrees of freedom, and the results of our algorithm can be sensitive to the extent of discretization error.
In addition, the user must decide on a trade-off between the number of collective coordinates used to discretize the state space, and the computational complexity. 
For example, our coordinate for the cones system lumps together the $T=4$ and $D5$ capsids, as these structures are very similar. While it would be straightforward to include an additional coordinate that distinguishes between the structures and then optimize specifically for $D5$ structures, we expect the resulting protocol would not change much because the $T=4$ yield is so low ($\approx5\%$ compared to $50\%$ for $D5$).
The size of the state space also affects computational and memory requirements. Machine learning approaches have the potential to address both these issues. Neural network architectures such as VAMPnets \cite{Mardt2018} or GraphVAMPnets \cite{Ghorbani2022} have been shown to streamline the MSM creation process by determining a minimal state space decomposition.

Finally, there is an implicit restriction on the temporal resolution of the protocols. The lower bound is set by the smallest lag time that results in convergence for all local MSMs. If this lag time is not sufficiently small compared to the assembly time-scales, the protocols are unlikely to be informative. This issue could be avoided in three ways: performing a finer discretization of state space such that a smaller lag time can be used; constructing a transition rate matrix \cite{Crommelin2009} and exponentiating it to construct the probability transition matrix, as we do for the colloids example; or formulating the optimization algorithm for a transition rate matrix directly, rather than a probability transition matrix.


\section*{Author Contributions}
\textbf{Michael Hagan}: Conceptualization (equal); Methodology (equal); Funding Acquisition (lead); Supervision (lead); Writing - Original Draft (equal); Writing - Review and Editing (equal).
\textbf{Anthony Trubiano}: Conceptualization (equal); Methodology (equal); Data Curation (lead); Formal Analysis (lead); Software (lead); Visualization (lead); Writing - Original Draft (equal); Writing - Review and Editing (equal). 

\begin{acknowledgments}
We would like to thank Eric Vanden-Eijnden for the idea for applying optimal control principles to self-assembling systems modeled by MSMs, and Miranda Holmes-Cerfon for helpful discussions on the method. 
We acknowledge support from NIH R01GM108021 and the Brandeis NSF MRSEC, Bioinspired Soft Materials, DMR-2011846. We also acknowledge computational support from NSF XSEDE computing resources allocation TG-MCB090163 and the Brandeis HPCC which is partially supported by the NSF through DMR-MRSEC 2011846 and OAC-1920147.
\end{acknowledgments}

\section*{Conflicts of Interest}
The authors declare no conflicts of interest. 

\section*{Data Availability Statement}

The code used to generate data, construct MSMs, perform optimization, and produce figures for the colloids model is publicly available at the Github repository: https://github.com/onehalfatsquared/CPfold. 

The code used to generate data, construct MSMs, perform optimization, and produce figures for the cones model is publicly available at the Github repository: https://github.com/onehalfatsquared/protocolOptMSM.

\section*{Supplementary Materials}
A supplemental PDF contains details on the implementation of simulation of the cones model, derivations for the adjoint method gradient calculation, details on performing gradient descent, details on MSM construction, details on the transition matrix interpolation and evaluation, and the full protocol optimization results for the colloidal polymer system.

\bibliographystyle{apsrev4-1}
\bibliography{references}


\appendix

\section{Simulation Details for Cones System}

Our simulations consist of two types of conical subunits \cite{Lazaro2018}, each made up of $6$ attractor beads, denoted $A_0$ through $A_5$, constrained to form a rigid body. We also include a spherical nanoparticle of radius $\Rnp$. 

The hexamer (H) is the reference subunit and is defined to have a cone height of $h=4\,$nm, cone angle of $\alpha = 17.8^{\circ}$, and outermost bead diameter of $\sigma_d = 7.7$nm. We can compute $\Rcone = \sigma_d / (2\sin(\alpha)) - h$ to be the distance between $A_0$ and an origin defined by the point of the cone with height $h$ and angle $\alpha$. $A_5$ is located at distance $\Rcone+h$, with the four interior beads equally spaced between the outer beads. The radii of each bead is then given by 
\begin{equation}
    r_i = (\Rcone + ih/5)\sin(\alpha), \quad i = 0, \dots, 5.
\end{equation}
The pentamers (P) are then constructed by multiplying the hexamer bead radii by a scaling factor $0.77$ and keeping the height constant. 
This scaling factor is roughly the ratio of the radii of the circle inscribed inside the pentagon to that of the hexagon in a chamferred dodecahedron \cite{GOLDBERG1937}.

There are two types of attractive interactions between the subunits, an H-H attraction and an H-P attraction, modeled via a Morse potential. Between two beads of type $A_i$, the bond has equilibrium length $\Lzero = 2r_i$, a cutoff distance of $\Lzero + 2$, a range parameter $\rho = 14 / \Lzero$, and well-depth as tunable parameters, $\EHH$ or $\EHP$. There is also an attractive Morse potential between $A_0$ and the spherical nanoparticle, with equilibrium length $\Lzero = r_0 + \Rnp$, cutoff distance $\Rnp+3$, range parameter $\rho=2/\Lzero$, and well-depths $\ENH = 7$ and $\ENP=6.3$. 

All other interactions are repulsive Lennard-Jones interactions. We impose excluded volume interactions between all pairs of unlike beads. We use a cutoff distance equal to the equilibrium length, $\Lzero=r_i+r_j$, for a bead of type $i$ interacting with type $j$. We use a well-depth $\epsilon_{ij} = 0.5$ for all combinations of $i\neq j$, except $\epsilon_{1,6} = 0.05$ to help with stability. We add longer range repulsive interactions between two pentamers beads of the same type by increasing the cutoff distance to $1.25\Lzero$, which helps promote assembly without defects. Finally, we impose excluded volume interactions between the nanoparticle and outermost beads $A_3$ through $A_5$ in the same way, with well-depth $\epsilon=0.5$. We allow overlap of the inner beads to help with stability of subunits bonded to the nanoparticle. 

The dimensionless distance in the simulation corresponds to $\lzero=1\,$nm, and energies are in units of $\kt$.
Interaction strengths are chosen in the range $1.2 \kt$ to $1.8 \kt$, which can facilitate assembly on the nanoparticle surface, but does not typically drive nucleation in the bulk. 
Each simulation uses $300$ capsomer subunits, split such that roughly $12/42$ are pentamers and $30/42$ are hexamers, and a single spherical nanoparticle with radius $\Rnp = 8.3\,$nm. 
The simulation domain is a cube with side length $120\,$nm. The time step is $0.02 \tzero$, and the systems are run for between $5\times10^6$ to $4\times10^7$ time steps, depending on the subunit-subunit interaction strengths, detailed in Section \ref{subsect:Sampling}. The simulations use the HOOMD-blue \cite{Anderson2020} Brownian Dynamics (BD) integrator, with an inverse temperature $\beta = 1$. 
The configurations are logged every $\Delta t = 25\tzero$ units of simulation time.

\section{Target Probability Gradient Derivation}  \label{sect:gradient}
Here we show the derivation of the probability gradient in Equation (4). We first need to express the adjoint vector, $\niceVec{F}{n}$, as a conditional expectation of the form
\begin{equation}
    F_i^n = E[V(X_N)|X_n=i],
\end{equation}
where $V$ is an arbitrary function, and $X_n$ is a random variable representing the state of the process at time step $n$. This can be shown to satisfy the Kolmogorov backward equation by utilizing the law of total probability and the Markov property, 
\begin{align}
    F_i^n &= E[V(X_N)|X_n=i] \\
    &= \sum_{j} E[V(X_N)|X_n = i, X_{n+1}=j] P(X_{n+1}=j|X_n=i) \\
    &= \sum_j P(X_{n+1}=j|X_n=i) E[V(X_N)|X_{n+1}=j] \\
    &= \mat{P}^n F_i^{n+1}. 
\end{align}
The final condition set for the backward equation corresponds to selecting $V(x) = \vec{1}_B(x)$, the indicator vector for the target set, $B$. Computing the expectation of $\vec{1}_B(X_N)$ gives the probability to maximize, $p_B^N$. Using the law of total probability, this is also equal to
\begin{align}
    p_B^N &= E[\vec{1}_B(X_N)] \\
    &= \sum_j E[\vec{1}_B(X_N)|X_i=j] P(X_i=j) \\
    &= \niceVec{p}{i} \cdot \niceVec{F}{i},
\end{align}
the inner product between the forward and backward equation solutions, evaluated at any intermediate time step, $i$.  Thus we can compute the relevant partial derivatives by a simple product rule to get
\begin{equation}\label{eq:prob_deriv_adjoint}
    \frac{\partial p_B^N}{\partial \theta_k} = \frac{\partial \niceVec{p}{i}}{\partial \theta_k} \niceVec{F}{i} + \niceVec{p}{i} \frac{\partial \niceVec{F}{i}}{\partial \theta_k}.
\end{equation}

This expression turns out to be quite useful, as any value of $i$ can be used in the computation. For a given $k$, an $i$ can be chosen such that one of the two derivatives on the right hand side is zero, and such that only a matrix-vector multiply is needed to evaluate the other derivative. For example, we know that $F^i$ is independent of $\theta_k$ if $i>k$, since the adjoint equation is propagated backwards in time. We can then choose $i=k+1$, making the second partial derivative zero. The expression becomes
\begin{align} 
    \frac{\partial p_B^N}{\partial \theta_k} &= \frac{\partial \niceVec{p}{k+1}}{\partial \theta_k} \niceVec{F}{k+1} + 0 \\
    &= \frac{\partial}{\partial \theta_k} \left(\niceVec{p}{0} \prod_{m=1}^{k} \mat{P}^m\right) \niceVec{F}{k+1} \\
    &= \niceVec{p}{k} \frac{\partial \mat{P}(\theta_k)}{\partial \theta_k} \niceVec{F}{k+1},
\end{align}
which only requires a matrix-vector product and an inner product to evaluate.

\section{Penalty Terms}
To ensure physically realizable protocols, we penalize variations in the parameters that occur too rapidly by adding the penalty term
\begin{equation}
    S[\theta] = -\frac{\lambda}{2}\sum_{j=0}^{N-1} \left(\frac{\theta_{j+1}-\theta_j}{\tau}\right)^2
\end{equation}
to the objective function. Directly taking the derivative of this penalty term with respect to the protocol at fixed times gives
\begin{equation} \label{eq:penalty_deriv}
    \frac{\partial S}{\partial \theta_k} = \frac{\lambda}{\tau^2} \left(\theta_{k+1} - 2\theta_k + \theta_{k-1}\right),
\end{equation}
which is proportional to a centered finite difference approximation of the second time derivative of the protocol.

We may also want to add another penalty term to favor faster assembly, or rather more stable assembly. One option is to include something like 
\begin{equation} \label{eq:stability_penalty}
    S_2[\theta] = -\frac{1}{2} \sum_{k=1}^N \mu_k (1-P_B^k)^2,
\end{equation}
which penalizes the target probability being different from $1$ at each time step. The $\mu_k$ can be chosen to be constant, or be time-dependent to reflect how important it is to be in the target at each time. The gradient of this penalty is
\begin{equation} \label{eq:stability_gradient}
    \frac{\partial S_2}{\partial \theta_j} = -\sum_{k=j+1}^N \mu_k (P_B^k-1)\frac{\partial P_B^k}{\partial \theta_j}.
\end{equation}
By solving the forward equation, we already know all the $P_B^k$. The partial derivatives are computed in exactly the same way as described previously, using the adjoint method, just with the terminal conditions set at each time step $k$ instead of $N$. Computationally, the forward equation only needs to be solved once, but the backward equation needs to be solved for each $k$, with terminal condition $\niceVec{F}{k} = \vec{1}_B$. We store the gradient of the probability for each $k$ using this method, noting that derivatives of $P_B^k$ with respect to $\theta_j$ are zero if $j > k$, i.e. if that value in the full protocol has not yet influenced the system at time $k$. 

Our code includes the option of adding this penalty function, but it was not used to generate the results shown here, as we found it was not necessary to achieve stable assembly.

\section{Gradient Descent} \label{sect:gd}

\subsection{Implementation}

Next, we describe how we perform gradient descent for the optimization. First, apply the adjoint-based gradient method to calculate the derivatives of the target probability with respect to each $\theta_k$. Let $\frac{\partial P_B^N}{\partial \theta_k} = G(N,k)$. Next, we combine this with the smoothness penalty term derivative and perform a gradient descent step with step size $h$ to update the protocol from iteration $n$ to $n+1$. This looks like
\begin{equation} \label{eq:gradient_descent_explicit}
\theta_{k}^{n+1} = \theta_k^{n} + hG^n(N,k) + \frac{\lambda h}{\tau^2} \left(\theta_{k+1}^n - 2\theta_k^n + \theta_{k-1}^n\right). 
\end{equation}
Note that this update looks like a forward Euler discretization of a heat equation with source term, e.g. $\theta_t(x,t) = G(x,t) + \lambda \theta_{xx}(x,t)$. From stability analysis of numerical PDEs, we know that this discretization would result in a severe restriction in the step $h$ due to the diffusive CFL condition for stability. With this in mind we instead discretize with an IMEX (Implicit-Explicit) method, where we treat the source term explicitly and the diffusive term implicitly. Doing so gives the linear system
\begin{equation} \label{eq:gradient_descent_linear_system}
\theta_{k}^{n+1} = \theta_k^{n} + hG^n(N,k) + \frac{\lambda h}{\tau^2} \left(\theta_{k+1}^{n+1} - 2\theta_k^{n+1} + \theta_{k-1}^{n+1}\right), 
\end{equation}
which can also be written in matrix form as
\begin{equation}
    (\mat{I}+\nu \mat{D})\niceVec{\theta}{n+1} = \niceVec{\theta}{n} + h\niceVec{G}{n},
\end{equation}
where $\nu = \lambda h /\tau^2$ is the diffusive CFL number, $\mat{D}$ is the centered second finite difference matrix, $\mat{I}$ is the $N \times N$ identity matrix, and $\niceVec{G}{n}$ is the probability gradient evaluated using the protocol at step $n$. This is a tri-diagonal system that can be solved quickly using the Thomas algorithm \cite{Press1992}. 

This system also needs to be equipped with two boundary conditions, due to the second derivative discretization. The choice of boundary conditions depends on the situation. If an experiment dictates specific initial and final values for the parameters, then Dirichlet conditions should be implemented. If the initial and final conditions can be chosen, then choosing Neumann boundary conditions will allow the system to determine the optimal start and end points, which is what we use in this paper. Combinations of the two types may also be used. 

Finally, note that the gradient of the $S_2$ penalty term does not explicitly depend on $\theta_j$. This means we can lump this gradient in with the probability gradient in the source term $G(N,k)$ and treat it explicitly, unlike the smoothing penalty term.

\subsection{Step Size and Convergence}

The value of $h$ at each time step can be set manually or determined by a line search. We initialize a starting value $h_0=1.5$, compute the new protocol according to the method above, and see if the result leads to an increase of the objective function. If not, we successively reduce the value of $h$ by a factor of two until we see an increase. Our convergence criteria is if the difference in successive objective function values is below a set tolerance of $10^{-5}$, or if the line search cannot find a step size that results in an increase of the objective function. 

We find that the solution determined by the gradient descent algorithm typically depends on the initial guess for the protocol, implying that the objective function landscape contains many local extrema. We implement a simple global optimization scheme to try to avoid these local extrema. To begin we initialize from a number of different initial conditions; various mixtures of constants, linear ramps, and oscillatory protocols. When the gradient descent is determined to have arrived at a critical point, we log the objective function value as well as the protocol. We then apply noise point-wise to the protocol and re-initialize the optimization. The noise is chosen to be either independent Gaussian noise with chosen variance, or a biased version where we ensure the noise is pointing in the direction of the last computed gradient. We keep track of the $10$ best critical points, stopping if four consecutive critical objective function values do not exceed the running maximum. This global optimization technique was applied to the cones system, but not the colloids system, where the optimization was more robust. This could reflect that the interpolated transition matrices for the cone system likely have noisier objective function landscapes than the semi-analytic model for the simpler colloidal polymer system.

\subsection{Effect of Regularization Parameter}

\begin{figure*}[t]
\centering  
\includegraphics[width=0.99\linewidth]{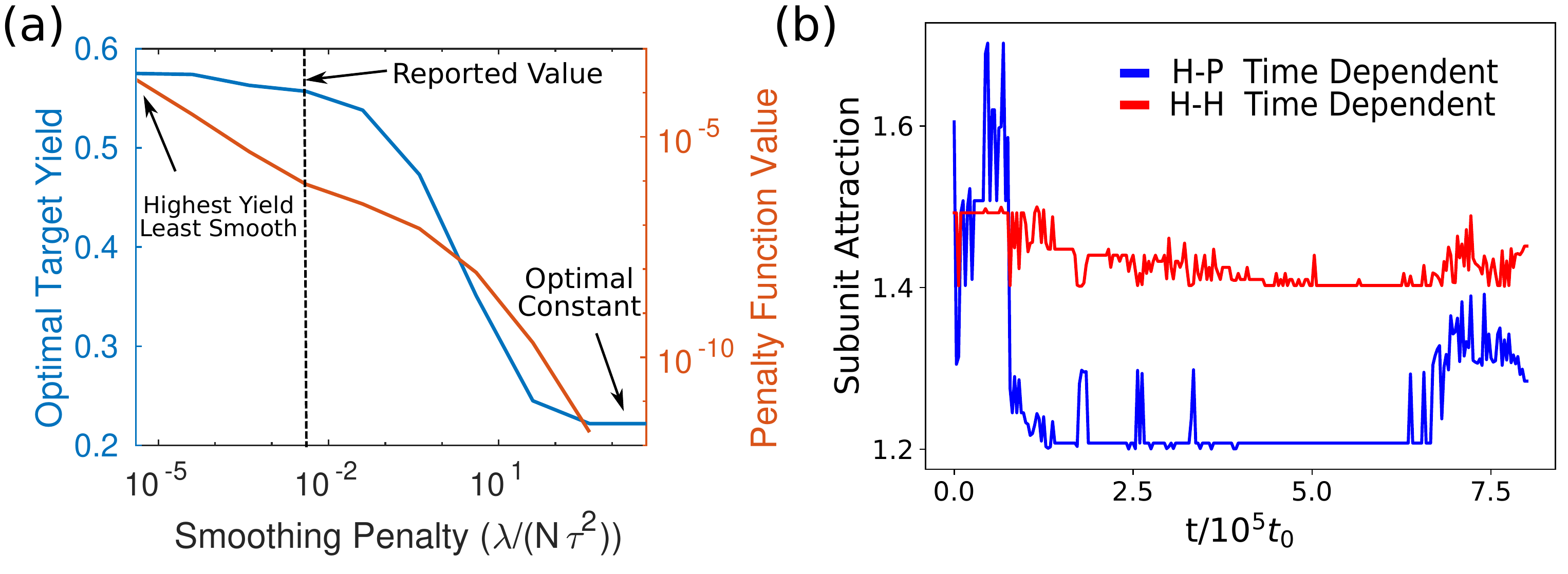}
\caption{Effect of regularization on optimal yield and protocol smoothness. (a) Plot of the optimal target yield in the cones system as a function of $\lambda^* = \lambda/(N\tau^2)$. Also plotted is the penalty function value for the resulting protocol. Large values correspond to large gradients in the protocol, and a value of zero corresponds to a constant protocol. The regularization used to generate the results in the main text is indicated by the dashed line. (b) Example protocol computed using the smallest regularization shown in (a), $\lambda^*=4\times 10^{-6}$. This protocol achieves the highest yield, but contains large and rapid oscillations. 
} 
\label{fig:regularization}
\end{figure*}

The value used for the regularization parameter, $\lambda$, controls the tradeoff between optimal target yield and smoothness of the resulting protocol. We report values for the re-scaled regularization parameter, $\lambda^* = \lambda/(N\tau^2)$, which normalizes the value with respect to the timescales of the particular system. 
Figure \ref{fig:regularization}(a) shows the optimal yield and protocol roughness as a function of $\lambda^*$ for the cones system optimization. We see that choosing a small value for the regularization results in the highest yield protocol, but also the least smooth one. This protocol is shown in Figure \ref{fig:regularization}(b), which is qualitatively similar to the optimal protocol shown in the main text, but with large and frequent oscillations in the values. 
As $\lambda^*$ increases over several orders of magnitude, the optimal yield is reduced as the protocol becomes smoother. Eventually, the penalty function becomes too strong and the protocol is reduced to the optimal constant values that were found by sampling values on an equally spaced grid with spacing $0.0075 \kt$ in both dimensions.  

\section{Markov State Models} \label{sect:MSM_app}

This section describes our procedure for constructing and validating the local Markov State Models for the cones system.

\subsection{State Space Discretization Coordinates}
First, we need a set of coordinates to describe the configurations in the simulations to generate discrete trajectories. 
We use the state space discretization coordinate $(\np, \nh, \IH)$ to define a state, where $\np$ and $\nh$ are the number of pentamers and hexamers attached to the nanoparticle, and $\IH$ is equal to the number of hexamers in contact with precisely two pentamers. Contacts are determined by setting cutoff distances. We use a cutoff distance of $13\,$nm between the nanoparticle center and the second outer-most cone bead to determine adsorption onto the nanoparticle. For subunit-subunit contacts, we use a cutoff distance $r_1 + r_2 + 2$, where $r_1$ and $r_2$ are the radii of the second outer-most bead of the corresponding subunits.   
The target state is specified by the triplet $(12,30,30)$, which corresponds to a $T=4$ or $D5$ capsid. 
We find that this three variable description is insufficient for a subset of configurations, in which case we augment our coordinate with the variables $\bhh$ and $\bhp$, the number of hexamer-hexamer and hexamer-pentamer bonds, respectively. As an example, the five coordinate state $(12,30,26,60,60)$ corresponds to a single, well-defined structure (structure C in Figure 1). The three coordinate description, $(12,30,26)$, lumps this state together with other transient structures, such as $(12,30,26,56,59)$, with distinct transition dynamics, so we refine our coordinate in this case. In total, our state space consists of $5727$ states. We discuss how we determine which states are insufficiently characterized by the three coordinate description after detailing how we perform our sampling of trajectories and construct MSMs.

\subsection{Sampling Trajectories} \label{subsect:Sampling}
Next we discuss how we sample trajectories to construct the MSMs. There are roughly three distinct regimes in the parameter space we are considering; 
a weak interaction regime where subunits will never nucleate on the nanoparticle ($1.2 \leq \EHP \leq 1.4$, $\EHH < 2.8 - \EHP$), 
an intermediate interaction regime where capsid growth occurs slowly but consistently for long times ($1.3 \leq \EHP \leq 1.5$, $\EHH \geq 2.8 - \EHP$),
and a strong interaction regime where capsid growth occurs quickly but typically with defects ($\EHP > 1.5$). 
We perform simulations differently in each of these regimes.  
In the intermediate interaction regime, we initialize $200$ trajectories with an empty nanoparticle and simulate until a final time of $\TF = 8\times10^5$. In the strong regime, the yield curves for the common states saturate much quicker, so we do the same but only simulate until $\TF = 5\times10^5$. The weak regime requires a different approach since initializing with an empty nanoparticle makes capsid growth an exceedingly rare event. Instead, we initialize the system in one of the five most common structures, shown in Figure 1, and simulate until $\TF = 8\times10^5$. We perform $40$ simulations in each of the five states as a baseline and add more as needed. In this parameter regime, the four off-target structures tend to transition to the ground state target structure, which remains stable, but occasionally the starting capsid will fully disassemble. 

In addition to these long simulations, we also perform shorter simulations that target rarely sampled transitions that are deemed important to the assembly process. To identify such states, we use the MSM to compute the average probability of each state over the simulation time, as well as the maximum probability of each state over the simulation. If the average probability of a state is greater than $10^{-5}$ or its max probability is greater than $0.005$, we identify that state as important and save a snapshot of the configuration. We perform this process for one parameter set in each of the three assembly regimes, and take the union of all important states. This leaves us with a total of $1497$ important states. For each of these states, we initialize three simulations in a snapshot of that state, and simulate until $\TF = 9.3\times 10^4$, or about $30$ lag times. 

\subsection{MSM Construction}

After trajectories have been sampled, they are converted to discrete trajectories via the state space discretization coordinate. Unique values of the state space discretization coordinate are assigned an integer index in the order they are first seen in the trajectories. 

The discrete trajectories are used to estimate count matrices at a lag time $\tau$. We use a sliding window, which counts transitions that occur between frame $t$ and frame $t+\tau$, for $t = 0, \cdots, T-\tau$. We use sparse matrix storage to keep track of all such transitions between pairs of states $i$ and $j$. For each MSM, the count matrix is of the same size, the total number of states observed among all parameter sets ($5727$ in our case), regardless of each state being observed for that particular parameter set or not. This means that some rows will be completely empty, depending on the parameter set used to generate the trajectories.

Next, we perform a pruning step. This eliminates entries corresponding to extremely rare and likely unimportant transitions and saves on memory. We set a cutoff of $2$ counts for self-transitions, i.e. the diagonal component of the count matrix. If a state is only observed for a single frame it is likely to be highly unstable and not worth tracking. We remove the corresponding row and column from the count matrix if it does not meet the cutoff. We also set a cutoff of $3$ counts for non-diagonal transitions. Transitions observed twice or less can be considered rare and are likely unimportant in characterizing the most common transition paths. We set counts that do not meet this cutoff to zero, removing them from the sparse matrix. 

We can now construct the probability transition matrix. This is done by summing each row of the count matrix to get a total number of counts, and then normalizing each row by this number. This is straight-forward, except in the case of a row with no observations. We refer to such states as \emph{inactive}. In this case, we simply put a $1$ on the diagonal to construct this row. During this process, we keep track of all such inactive states. We can then construct a reduced transition matrix, only over the active states, by removing the rows and columns of inactive states, and creating a mapping between state indices in the full and reduced forms. We use the reduced transition matrices for all calculations. Finally,  we do not impose reversibility during MSM construction, since capsid assembly is out of equilibrium and thus the sampled trajectories are not reversible (although the underlying dynamics obeys microscopic reversibility).  It should be possible to achieve reversible sampling by combining the unbiased assembly simulations with free energy calculations or other enhanced sampling methodologies to sample improbable disassembly transitions, but that is beyond the scope of the present study since the unbiased trajectories enable estimating a sufficiently accurate MSM.

\subsection{Convergence and Validation}

\begin{figure*}[t]
\centering  
\includegraphics[width=0.99\linewidth]{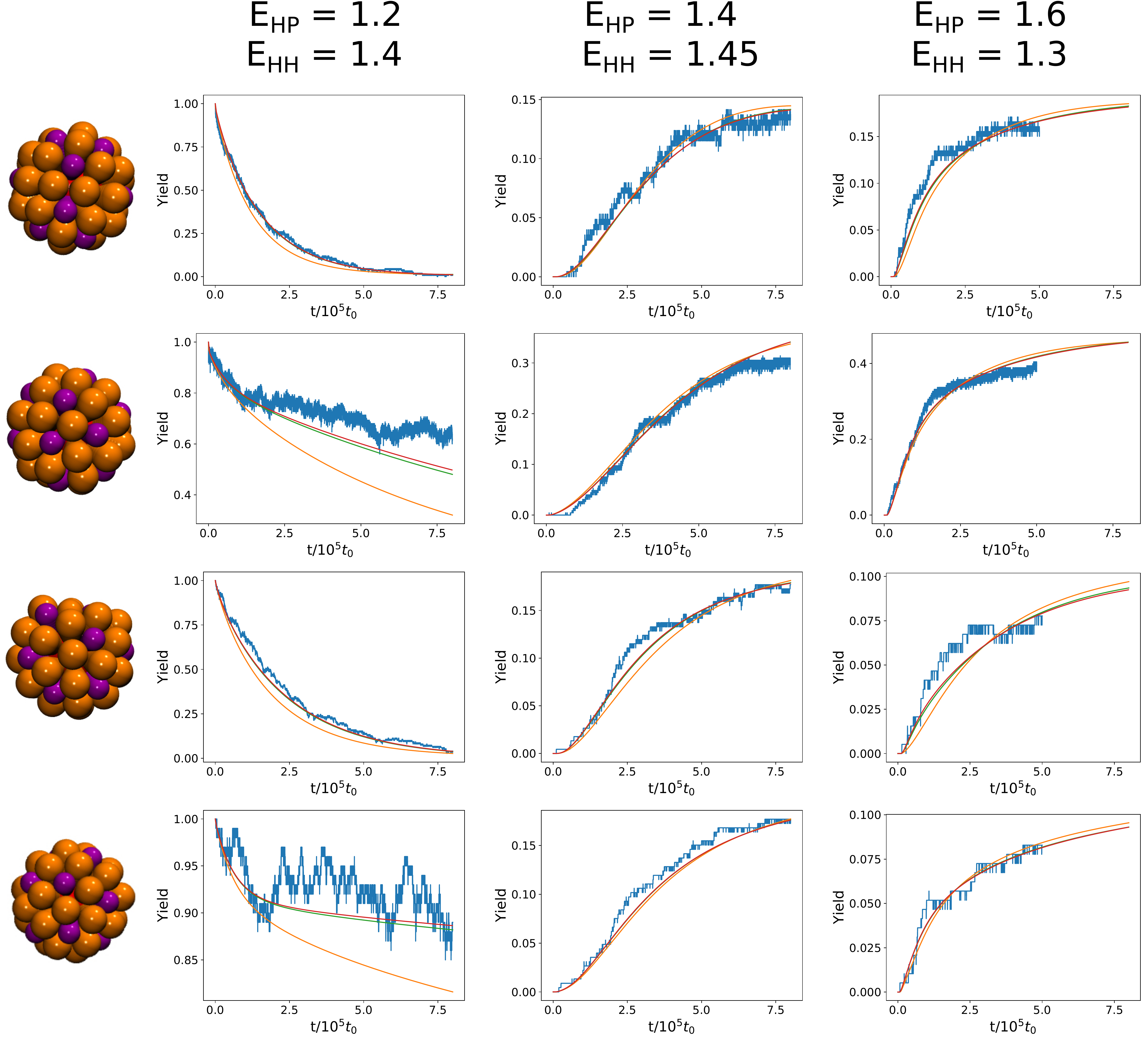}
\caption{Yield curves for the four most common cone assemblies for various values of the MSM lag-time, $\tau$. Tests are performed for a single parameter set in each of the three assembly regimes; weak interactions (left), intermediate interactions (middle), and strong interactions (right). Blue curves are estimated directly from sampled trajectories. The other curves are computed from an MSM with lag-times $625\tzero$ (orange), $3125\tzero$ (green), and $3750\tzero$ (red). Note that red and green curves are essentially overlapping in each plot, signifying convergence with respect to the choice of lag-time. 
} 
\label{fig:verification}
\end{figure*}

MSMs can describe system dynamics at long timescales with high accuracy, but this requires a suitable choice of discretization and lag time \cite{Prinz2011}. The typical approach to testing convergence of an MSM is to perform a Chapman-Kolmogorov (CK) test or to compute the largest implied timescales from the transition matrix and check if they are lag-time-independent near the chosen value of $\tau$. These tests are difficult to apply to our system. We find unconverged timescales no matter how large we choose $\tau$, but these correspond to absorbing or nearly-absorbing states with many defects that are unimportant to the capsid assembly process. Based on this, we perform a validation that is better suited to and more stringently tests the assembly problem, used here \cite{Perkett2014}. 

To gauge convergence of the MSMs, we compute yield curves for the four most common assembled capsids for various lag-times. These yield curves track the fraction of trajectories in a particular state as a function of time. These curves will depend on all the implied timescales that are relevant to that particular assembly process, not just the longest timescales, so this is a stricter measure of the MSM's convergence. Figure \ref{fig:verification} shows these yield curves for the four most common structures at an example parameter set in each of the three regimes of interaction strengths. Note that the left column tracks disassembly starting from the shown state, while the other plots track assembly to that state starting from an empty nanoparticle. We find that the yield curves using lag-times of $3125\tzero$ and $3750\tzero$ are essentially overlapping in each plot, which validates the construction of MSMs using $\tau=3125\tzero$ in our reported results. In addition to testing convergence, we also compare the MSM predicted yield curves to direct estimates from sampling. The agreement is excellent for most cases, and at least qualitatively accurate on average in the worst cases. 

\subsection{State Space Refinement}

We now describe how we choose which states in our discretization to refine, from the three coordinate description, $(\np, \nh, \IH)$, to the five coordinate description, $(\np, \nh, \IH, \bhh, \bhp)$. 

We begin by constructing an MSM using half of the available simulation trajectories, split randomly. For each state, we can access the corresponding row in the count matrix as well as the transition matrix. This allows us to estimate a standard error for our transition matrix estimates. By treating each row of the count matrix as a multinomial distribution, where the states are bins, the standard error of bin $i$, $\se(i)$, can be computed as 
\begin{equation}
    \se(i) = \sqrt{\frac{p_i(1-p_i)}{N}}
\end{equation}
where $p_i$ is the transition probability of state $i$ and $N$ is the total number of counts in the row. 

Once we have these standard errors for each row, we can compute another MSM using the full set of simulation data and compare these new estimates, $p_i'$, to the base estimates, $p_i$. If the new estimates differ from the base estimates by some tolerance, we conclude that the samples are likely sampling different regions of the state space, so the state corresponding to row $i$ should be refined. We set the tolerance to $|p_i-p_i'| > 3\se(i)$.

\section{Interpolation} \label{sect:interpolation}

\subsection{Radial Basis Function Background}

Consider the interpolation problem where we are given $N$ points, $\{x_i\}_{i=0}^{N-1}$, $x_i\in \mathbb{R}^d$, as well as $N$ values $\{y_i\}_{i=0}^{N-1}$ such that each of these values is assumed to be sampled from some unknown function, i.e. $f(x_i) = y_i$. The goal is to construct a function that satisfies the known data, but can be extrapolated to other values of the input. 

The radial basis function (RBF) method is a mesh free way of constructing such an interpolating function. Given the data, which is potentially sampled in an unstructured way, an interpolant is constructed as a linear combination of a set of radial basis functions, or kernels, $\phi(r)$. The expression for the interpolant is given by
\begin{equation} \label{eq:rbf}
    \hat{f}(x) = \sum_{i=0}^{N-1} c_i \phi(\|x-x_i\|_2),
\end{equation}
where the $c_i$ are constant coefficients chosen to satisfy the $N$ constraints from the observed data. This gives a linear system of the form $Kc=y$ to solve for the coefficients, where $K_{ij} = \phi(\|x_i-x_j\|_2)$. Since this system is linear and symmetric, a unique solution to the interpolation will exist if the matrix is also positive definite. Certain kernels guarantee positive definiteness of the matrix $K$, but numerical issues such as round-off errors can introduce small, negative eigenvalues that may prevent a solution from being found. 

Here we will use the Gaussian kernel, $\phi(r) = e^{-\epsilon r^2}$. Note that this expression contains a parameter, $\epsilon$, called the shape parameter, that affects the width of the basis functions as well as how accurate the resulting interpolant is. Smaller values of the shape parameter result in flatter basis functions, which spread the information from the given data points further away from the source nodes. Typically, the interpolation error decreases with smaller shape parameter, up to a point, and then increases again beyond this value. Thus, constructing an RBF interpolant will involve a procedure for determining what value of the shape parameter minimizes the interpolation error. 

A simple method for determining the optimal value for the shape parameter is known as cross-validation. More specifically, the leave-one-out cross validation (LOOCV) is well suited to this problem. Here, the interpolant is constructed on $N-1$ data points, and then the final data point is used to get a measure of the error. This can be repeated for each point, giving a vector of $N$ errors whose $l^2$ norm can then be minimized to determine an optimal shape parameter. 

The direct method as stated above is quite inefficient; each linear solve to construct the RBF interpolant involves $O(N^3)$ computation, which must be done $N$ times, giving an $O(N^4)$ algorithm for each choice of the shape parameter. This can be reduced to $O(N^3)$ by the use of an interesting property of the error. It can be shown \cite{Rippa1999} that the error incurred by excluding the $k$-th data point can be computed as 
\begin{equation} \label{eq:Rippa}
    e_k = \hat{f}_k(x_k)-y_k = \frac{c_k}{(K^{-1})_{kk}},
\end{equation}
which only involves inversion of a single linear system, the system required to construct the interpolant in the first place. With this result, an expression for the error comes for free with computation of the interpolant. We identify that the error minimizing shape parameter typically takes values in the interval $[0.1,100]$ for our problem. To choose $\epsilon$, we sample $200$ logarithmically spaced points in this interval, compute the errors according to the above formula, and determine which value globally minimizes the $l^2$ norm of the error.

\subsection{Constructing Interpolants}

The first step in constructing the transition matrix interpolant is to define a global state space. Each locally estimated MSM has its own set of active states, which we will call the active set. We take the union of all of these active sets to construct the global active set. 
We also determine a mapping that converts a state index from the original full state space to an index in the global active set and vice-versa. 
The nodal points for the interpolation are all parameter values, $\vec{E}_k = (\EHP^k, \EHH^k)$, in which MSMs are estimated, which we represent as $x=\{\vec{E}_k\}_{k=1}^K$, where $K$ is the number of parameter values sampled. We then consider each pair of states, $i$ and $j$, that have a nonzero transition probability in \emph{any} of the $K$ local MSMs. We construct a list of values $y = \{P_{ij}(\vec{E_k})\}_{k=1}^K$, the estimated transition probabilities for that transition in each MSM. 
If a particular transition is observed for some $k$, but not all, then we append a transition probability of $0$ when it is not observed.
If a transition is not observed in any of the $K$ MSMs, we leave this entry empty in the sparse matrix storing the values. 

Once the sparse matrix of transition probabilities has been filled, we pass each entry to the interpolation scheme that computes the optimal shape parameter, using the above algorithm, and returns the interpolant object. Each entry is independent, so this calculation can be trivially parallelized. The resulting interpolants can be stored in a sparse matrix and then serialized for future access. If the system becomes too large, it may become difficult to store all interpolants in RAM, as becomes the case with our system with $5727$ states. For this case, we construct an SQL database on the HDD using SQLite3 \cite{Hipp2020} to store interpolants. 
We use a python package \cite{Hines2016} for computing the radial basis function interpolants.

\subsection{Evaluating Interpolants}

To evaluate a transition matrix at a desired valued of the parameters, $\theta$, we evaluate each of the stored interpolants at that value, which consists of plugging in the coefficients we solve for into Equation \eqref{eq:rbf}, and organizing them into a sparse matrix. Before using this transition matrix, we perform a re-normalization procedure on each of the rows. This serves two purposes; to ensure we use a valid transition matrix in the optimization and to reduce the number of entries for both accuracy of the model and memory savings. Interpolation typically results in many small non-zero entries, on the order of $10^{-5}$, where the nodal values should be $0$. Over long timescales, such as the assembly timescale we consider, this can result in spurious dynamics such as leaking probability out of an absorbing state, so we prune these values. Further, noise induced by interpolation error results in un-normalized probability distributions that must be normalized before a valid MSM can be used for optimization. 

Our normalization procedure for the $j$-th row of the transition matrix, denoted $\vec{p}^j(\theta)$, is as follows. First we compute the probability row sum, $S^j(\theta) = \sum_i p_i^j(\theta)$, and check if this is greater than a threshold of $p_{\text{cut}} = 0.02$. If not, we put a $1$ on the diagonal entry for the row and go to the next. 
If the first threshold is met, we let $M$ denote the largest entry in the row. We then check if each element is greater than $M/R$, where $R=40000$ is a reference scaling constant determined by trial and error. If an entry is less than $M/R$, we append it to a pruned set, $P^j$, and add the value to a running sum, $S_P^j(\theta)$. We then set the entry to zero. Otherwise, we keep the entry and add the index to the kept set, $K^j$. 
We then distribute the total pruned probability, $S_P^j(\theta)$, equally to each non-zero entry in the row except the state on the diagonal as to not bias towards self-loops. Finally, we divide each entry by the row sum to ensure the distribution is normalized. Denoting the normalized row $\vec{n}^j(\theta)$, it can be computed as 
\begin{equation}
    n_i^j(\theta) = \frac{u_i^j(\theta)}{S^j(\theta)} = 
        \frac{1}{S^j(\theta)} 
        \begin{cases}
            p_i^j(\theta) & \text{if } i = j,\\
            p_i^j(\theta) + \frac{1}{|K|}S_P^j(\theta) & \text{if } i \in K^j \text{ and } i \neq j,\\
            0 & \text{if } i \in P^j,
        \end{cases}
\end{equation}
where $u_i^j(\theta)$ is the un-normalized probability after the pruning step. 
This procedure has the effect of removing entries that are not sufficiently large compared to the largest probability in the row. This is intended to keep the important probabilities, while suppressing the small, spurious entries that come from interpolating values near $0$.  

We also need to compute partial derivatives of the transition matrix with respect to each parameter for the optimization. The evaluation of derivatives of $u_i^j(\theta)$ is trivial; one only needs to differentiate the basis functions, which are Gaussian in this case. The tricky step is accounting for the pruning on elements and re-normalization of rows in the transition matrix when computing the derivative matrix. Applying the quotient rule to the expression for the normalized probability, we get
\begin{align}
    d_i^j(\theta) &= \frac{(\partial_\theta u_i^j(\theta))S^j(\theta) - u_i^j(\theta)(\partial_\theta S^j(\theta))}{(S^j(\theta))^2}\\
    &= \frac{1}{S^j(\theta)} \left( \partial_\theta u_i^j(\theta) - n_i^j(\theta) \partial_\theta S^j(\theta)\right)
\end{align}
as an expression for the rows of the derivative of the transition matrix. 

Evaluating an individual interpolant is computationally simple, but evaluating a large number of them when the system size gets large can be demanding. For this reason, we do not always evaluate the transition matrices on the fly. Instead, we set up a discretization and caching system. 
We discretize our feasible parameter region on a uniform grid with spacing $0.0075$ between grid points. When an interpolant is requested at parameter $\theta$, we determine the grid point closest to $\theta$ and evaluate the interpolant there. After an interpolant and its derivatives have been evaluated at a grid point, we cache the resulting transition matrices into a dictionary stored on the HDD, using the python \emph{shelve} module. The next time a transition matrix is requested at this grid point, we retrieve it from the dictionary instead of computing it from scratch.

\section{Detailed Results for Colloid Optimization}

\begin{figure*}[t]
\centering  
\includegraphics[width=0.99\linewidth]{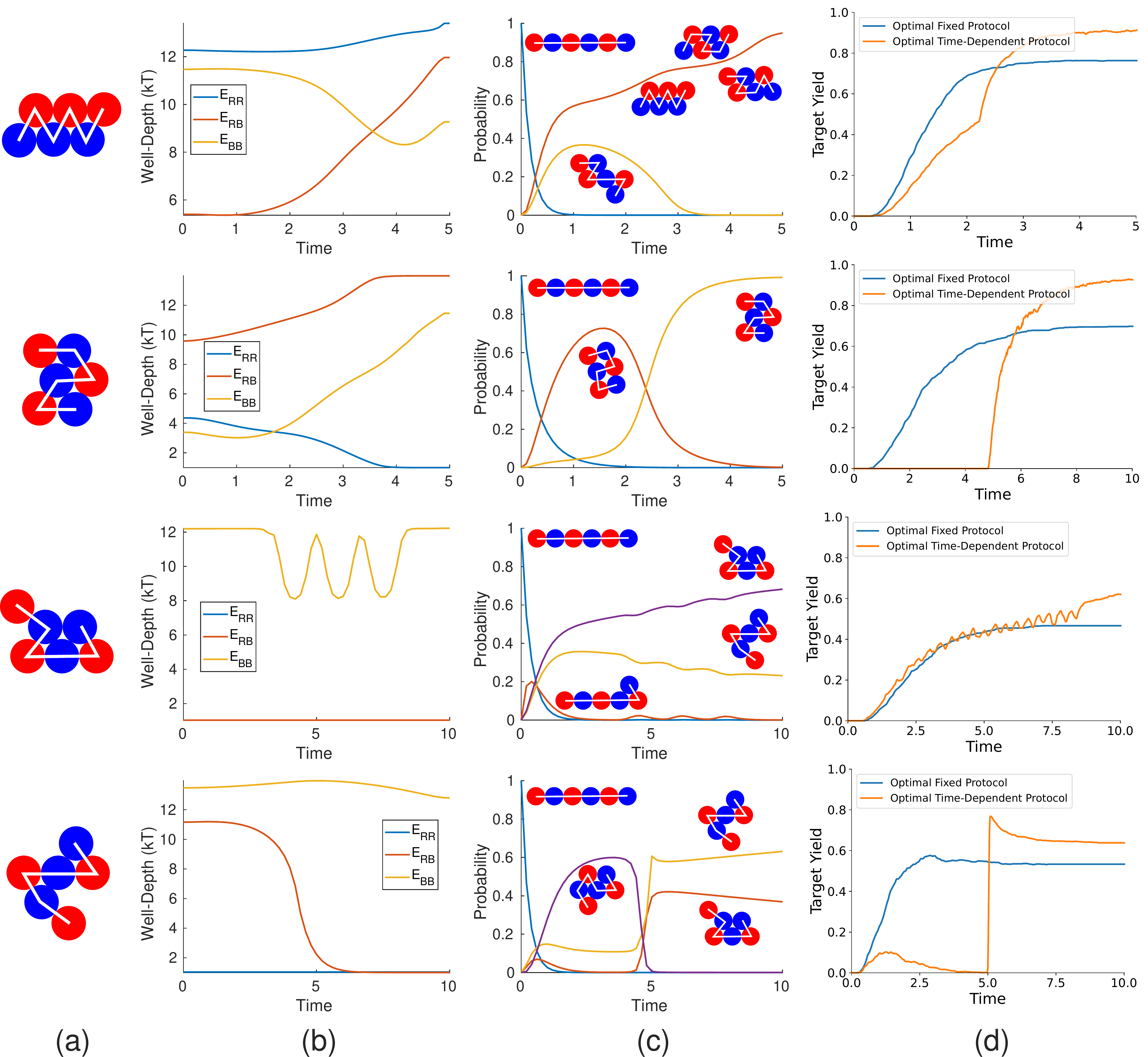}
\caption{Protocol optimization results and verification for the colloidal polymer system. 
(a) The structure being optimized for in each row. Note that only a single permutation is shown, but all permutations consistent with the target structure are lumped for the optimization. 
(b) The protocols for each interaction strength determined by the optimization. 
(c) Probabilities as a function of time for all structures achieving a maximum yield of over $0.2$ using the optimal protocol in (b), evaluated using the MSM. 
(d) Time-dependent yield of the target structure averaged over $600$ Brownian dynamics trajectories using both the optimal fixed interactions and the optimal protocol in (b). 
} 
\label{fig:colloidOptSM}
\end{figure*}

The optimization results and verification for each of the colloidal structures in the main text are summarized in Figure \ref{fig:colloidOptSM}. 
The parallelogram and chevron, structures $1$ and $2$, respectively, were optimized starting from an initial guess of a constant energy of $6\, \kt$ for each interaction. Both structures see yields increase from around $70\,\%$ with fixed interactions, to over $90\,\%$ with a time-dependent protocol.
The optimal protocol for the parallelogram begins with large $\ERR$ and $\EBB$, and weak $\ERB$. This promotes assembly of the permutation shown in Figure \ref{fig:colloidOptSM}(a). It then slightly lowers $\EBB$ but such that the assembled permutation remains stable, while increasing $\ERB$, which promotes assembly of other permutations that are stuck in floppy configurations. 
The chevron protocol begins with just a large $\ERB$, which selectively forms a structure similar to the chevron, but with the middle row of particles aligned with the other to form a rectangle. By switching on $\EBB$, the chevron becomes the only possible structure to form.

The floppy structures, $4$ and $5$, both form with about $50\%$ probability when only $\EBB$ is allowed to be non-zero and it is chosen to be large. Structure $4$ is selected when all blue-blue bonds form on the same side of the central blue particle, whereas structure $5$ forms when the blue-blue bonds are on opposite sides of the central blue particle. 

There is a degeneracy issue in the Markov model that must be addressed before optimizing for these structures. It is clear from looking at the bond networks that structure $5$ cannot form structure $4$ without first breaking bonds. However, the adjacency matrix for structure $5$ is identical to that of a structure made by taking structure $4$ and removing the bond between the upper blue particles. This structure \emph{can} transition to structure $5$, and this degeneracy in the adjacency matrix representation leads to the MSM predicting structure $4$ forming with $100\,\%$ yield with only a blue-blue interaction, when really it should be split roughly equally between these two structures. 

To fix this issue we manually modify a few of the estimated transition rates between adjacency matrices. To start, we set the transition rate of structure $5$ to structure $4$ to zero. Let structure $S_{24}$ denote the structure where there is one bond between particles $2$ and $4$, and $S_{46}$ denote the structure with a bond between particles $4$ and $6$.  We then make the approximation that if either of these structures form structure $4$, both of the additional bonds form at once. We further approximate that these transitions happen at the same rate of both of these structures transitioning to structure $5$, which we already have estimates for. The resulting MSM gives then gives approximately $50/50$ odds of forming structures $4$ and $5$ when only the blue-blue interaction is active. 

By performing the optimization with this modified transition matrix, the yields of these structures can be slightly boosted to around $63 \%$ using different protocols. 
Structure $4$ has three blue-blue bonds, making it more stable than structure $5$. It follows then that a tempering protocol that increases and decreases $\EBB$ would promote transitions from structure $5$ to structure $4$, while leaving the already assembled structure $4$s intact. We use an initial guess of zero for the red-red and red-blue interactions strengths, and a cosine protocol oscillating between $6 k_BT$ and $14 k_BT$ over six periods. The resulting protocol begins and ends with a strong blue-blue interaction, with oscillations in between, which does indeed boost the yield of structure $4$. We note that in order to observe oscillations in the optimization result, they needed to be present in the initial guess. This is likely due to the smoothing parameter penalizing the introduction of too much variation in the protocol. 

To optimize for structure $5$, we used constant initial guesses of $6 k_BT$ for each pair of interactions, while leaving the third near zero.  
The final protocol for structure $5$ is relatively simple; the yield can be boosted by adding in a red-blue interaction initially, then turning it off midway through the assembly.

\end{document}